\documentclass[conference]{IEEEtran}
\IEEEoverridecommandlockouts
\usepackage{cite}
\usepackage{amsmath,amssymb,amsfonts}
\usepackage{algorithmic}
\usepackage{algorithm}
\usepackage{graphicx}
\usepackage{tabularx}
\usepackage{textcomp}
\usepackage{xcolor}
\def\BibTeX{{\rm B\kern-.05em{\sc i\kern-.025em b}\kern-.08em
    T\kern-.1667em\lower.7ex\hbox{E}\kern-.125emX}}
    
\usepackage{microtype}
\usepackage{graphicx}
\usepackage{subfigure}
\usepackage{booktabs} % for professional tables
\usepackage{multirow}
\usepackage{acronym}
\usepackage{tabu}
\usepackage{cite}
\usepackage{amsmath}
\usepackage{lettrine}
\usepackage{tikz}

\usepackage{orcidlink}
\usepackage[flushleft]{threeparttable}
\acrodef{DAC}[DAC]{Digital-to-Analog Converter}
\acrodef{ADC}[ADC]{Analog-to-Digital Converter}
\acrodef{DNN}[DNN]{Deep Neural Network}
\acrodef{NAS}[NAS]{Neural Architecture Search}
\acrodef{HWA}[HWA]{Hardware-Aware}
\acrodef{HW-NAS}[HW-NAS]{Hardware-Aware Neural Architecture Search}
\acrodef{DL}[DL]{Deep Learning}
\acrodef{MAC}[MAC]{Multiply-Accumulate}
\acrodef{IMC}[IMC]{In-Memory Computing}
\acrodef{PCM}[PCM]{Phase Change Memory}
\acrodef{ML}[ML]{Machine Learning}
\acrodef{SGD}[SGD]{Stochastic Gradient Descent}
\acrodef{CE}[CE]{Cross-Entropy}
\acrodef{SOTA}[SOTA]{State-Of-The-Art}
\acrodef{MVM}[MVM]{Matrix-Vector Mutliplication}
\acrodef{LHS}[LHS]{Latin Hypercube Sampling}
\acrodef{MSE}[MSE]{Mean Squared Error}
\acrodef{VWW}[VWW]{Visual Wake Words}
\acrodef{KWS}[KWS]{Keyword Spotting}
\acrodef{RGB}[RGB]{Red Green Blue}
\acrodef{CNN}[CNN]{Convolutional Neural Network}
\acrodef{FPGA}[FPGA]{Field-programmable Gate Array}
\acrodef{GPGPU}[GPGPU]{General-purpose Graphics Processing Units}
\acrodef{ASIC}[ASIC]{Application-specific Integrated Circuit}
\acrodef{RRAM}[RRAM]{Resistive Random Access Memory}
\acrodef{MRAM}[MRAM]{Magnetic Random Access Memory}
\acrodef{PCM}[PCM]{Phase Change Memory}
\acrodef{TinyML}[TinyML]{Tiny Machine Learning}
\acrodef{CPU}[CPU]{Central Processing Unit}
\acrodef{GPU}[GPU]{Graphics Processing Unit}
\acrodef{AIHWKIT}[\textsc{AIHWKit}]{IBM Analog Hardware Acceleration Kit}
\acrodef{IoT}[IoT]{Internet of Things}
\acrodef{NVM}[NVM]{Non-Volatile Memory}
\acrodef{WL}[WL]{Word-Line}
\acrodef{HWA}[HWA]{Hardware-Aware}
\acrodef{AVM}[AVM]{Accuracy Variation over One Month}

\author{\IEEEauthorblockN{Hadjer Benmeziane\IEEEauthorrefmark{1}\orcidlink{0000-0002-5259-0749},
Corey Lammie\IEEEauthorrefmark{2}\orcidlink{0000-0001-5564-1356},
Irem Boybat\IEEEauthorrefmark{2}\orcidlink{0000-0002-4255-8622}, 
Malte Rasch\IEEEauthorrefmark{3}\orcidlink{0000-0002-7988-4624},Manuel Le Gallo\IEEEauthorrefmark{2}\orcidlink{ 0000-0003-1600-6151},  Hsinyu Tsai\IEEEauthorrefmark{4}\orcidlink{0000-0002-3971-097X}, \\Ramachandran Muralidhar\IEEEauthorrefmark{3}\orcidlink{0000-0002-3982-3288}, Smail Niar\IEEEauthorrefmark{1}\orcidlink{0000-0002-7550-484X}, Ouarnoughi Hamza\IEEEauthorrefmark{1}\orcidlink{0000-0002-7490-5350}, Vijay Narayanan\IEEEauthorrefmark{3}, \\ Abu Sebastian\IEEEauthorrefmark{2}\orcidlink{ 0000-0001-5603-5243} and Kaoutar El Maghraoui\IEEEauthorrefmark{3}\orcidlink{ 0000-0002-1967-8749},}
\IEEEauthorblockA{\IEEEauthorrefmark{1}Univ. Polytechnique Hauts-de-France, CNRS, UMR 8201 - LAMIH, F-59313 Valenciennes, France}
\IEEEauthorblockA{\IEEEauthorrefmark{2}IBM Research Europe, 8803 Rüschlikon, Switzerland}
\IEEEauthorblockA{\IEEEauthorrefmark{3}IBM T. J. Watson Research Center, Yorktown Heights, NY 10598, USA}
\IEEEauthorblockA{\IEEEauthorrefmark{4}	IBM Research Almaden, 650 Harry Road, San Jose, CA USA}% <-this % stops an unwanted space
\thanks{\hspace{-1em}\rule{3cm}{0.5pt} \newline \textcopyright  \hspace{1pt} 2023 IEEE. Personal use of this material is permitted. Permission from IEEE must be obtained for all other uses, in any current or future media, including reprinting/republishing this material for advertising or promotional purposes, creating new collective works, for resale or redistribution to servers or lists, or reuse of any copyrighted component of this work in other works.}
}
\begin{document}

%\title{Analog-NAS: Neural Architecture Search \\ for Analog AI}
\title{AnalogNAS: A Neural Network Design Framework for Accurate Inference with Analog In-Memory Computing}

\maketitle

\begin{abstract}
The advancement of \ac{DL} is driven by efficient \ac{DNN} design and new hardware accelerators. Current \ac{DNN} design is primarily tailored for general-purpose use and deployment on commercially viable platforms. Inference at the edge requires low latency, compact and power-efficient models, and must be cost-effective. Digital processors based on typical von Neumann architectures are not conducive to edge AI given the large amounts of required data movement in and out of memory. Conversely, analog/mixed-signal in-memory computing hardware accelerators can easily transcend the memory wall of von Neuman architectures when accelerating inference workloads. They offer increased area- and power efficiency, which are paramount in edge resource-constrained environments. 
%Analog hardware is rarely considered when designing \acp{DNN}. %\textcolor{red}{Maybe we can talk about HW-NAS?} 
In this paper, we propose \textit{AnalogNAS}, a framework for automated \ac{DNN} design targeting deployment on analog \ac{IMC} inference accelerators. We conduct extensive hardware simulations to demonstrate the performance of AnalogNAS on \ac{SOTA} models in terms of accuracy and deployment efficiency on various \ac{TinyML} tasks. We also present experimental results that show AnalogNAS models achieving higher accuracy than \ac{SOTA} models when implemented on a 64-core \ac{IMC} chip based on \ac{PCM}. The AnalogNAS search code is released\footnote{https://github.com/IBM/analog-nas}.
\end{abstract}


\section{Introduction}
\label{intro}
\lettrine[lines=2]{W}{ith} the growing demands of real-time \ac{DL} workloads, today's conventional cloud-based AI deployment approaches do not meet the ever-increasing bandwidth, real-time, and low-latency requirements. Edge computing brings storage and local computations closer to the data sources produced by the sheer amount of \ac{IoT} objects, without overloading network and cloud resources. As \acp{DNN} are becoming more memory and compute intensive, edge AI deployments on resource-constrained devices pose significant challenges. These challenges have driven the need for specialized hardware accelerators for on-device \ac{ML} and a plethora of tools and solutions targeting the development and deployment of power-efficient edge AI solutions. %\textcolor{red}{refs?} If enough space we can add ~\cite{DBLP:journals/corr/abs-1906-03138}
One such promising technology for edge hardware accelerators is analog-based \ac{IMC}, which is herein referred to as \textit{analog \ac{IMC}}.

Analog \ac{IMC}~\cite{imcgeneral} can provide radical improvements in performance and power efficiency, by leveraging the physical properties of memory devices to perform computation and storage at the same physical location. Many types of memory devices, including Flash memory, \ac{PCM}, and \ac{RRAM}, can be used for \ac{IMC}~\cite{DBLP:journals/corr/abs-1906-03138}. Most notably, analog \ac{IMC} can be used to perform \ac{MVM} operations in $O(1)$ time complexity~\cite{Lammie2022}, which is the most dominant operation used for \ac{DNN} acceleration. In this novel approach, the weights of linear, convolutional, and recurrent DNN layers are mapped to crossbar arrays (tiles) of \ac{NVM} elements. By exploiting basic Kirchhoff's circuit laws, \acp{MVM} can be performed by encoding inputs as \ac{WL} voltages and weights as device conductances. For most computations, this removes the need to pass data back and forth between \acp{CPU} and memory. This back and forth data movement is inherent in traditional digital computing architectures, and is often referred to as the \textit{von Neumann bottleneck}. Because there is greatly reduced movement of data, tasks can be performed in a fraction of the time, and with much less energy.

\begin{figure}[!t]
    \centering
    \includegraphics[width=0.45\textwidth]{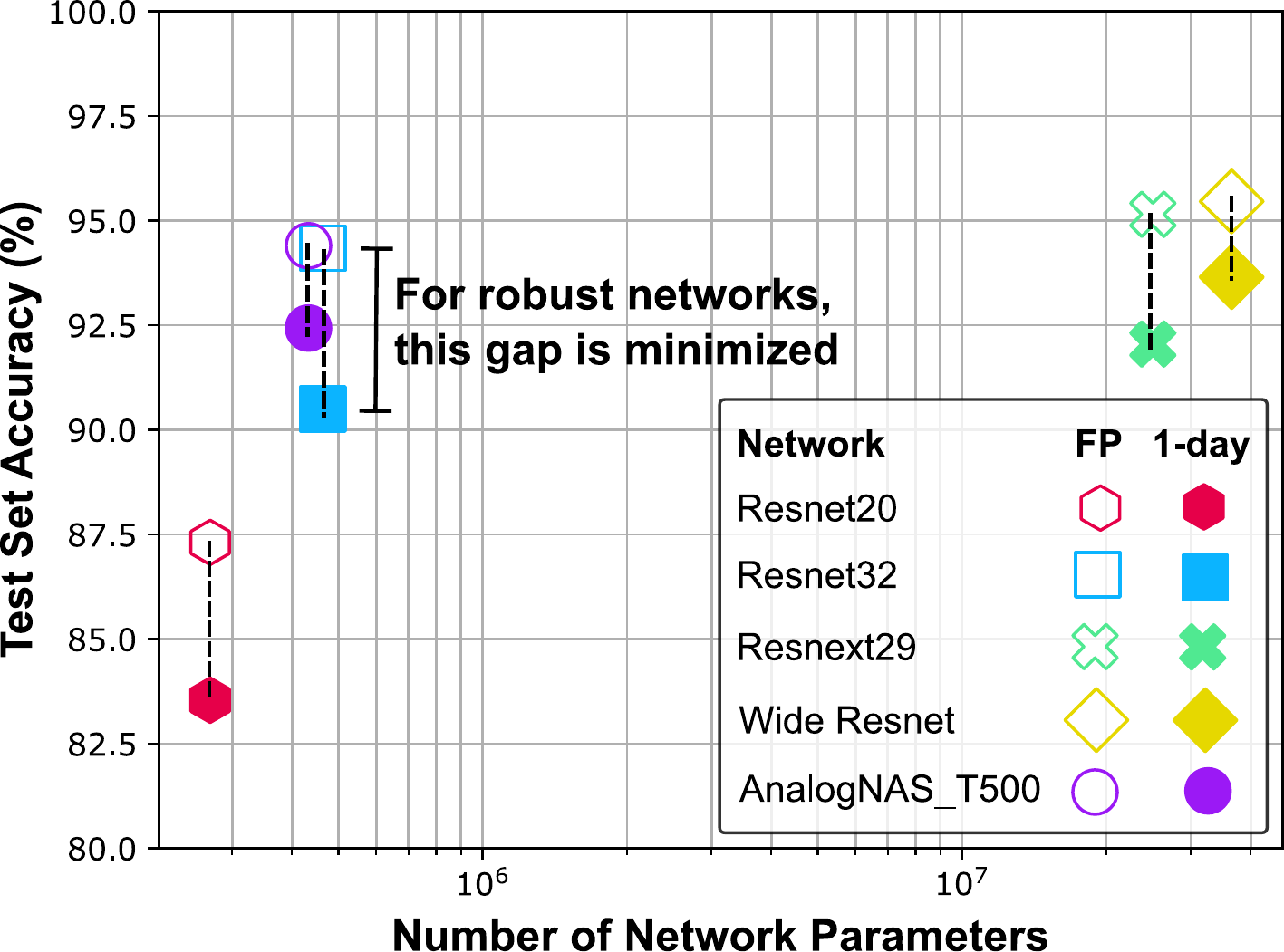}
    \caption{The effect of \ac{PCM} conductance drift after one day on standard \ac{CNN} architectures and one architecture (\texttt{AnalogNAS\_T500}) obtained using \ac{HW-NAS}, evaluated using CIFAR-10. \textit{FP} refers to the original network accuracy, and \textit{1-day} to the simulated analog network accuracy after 1-day device drift. }
    \label{fig:motivation}
\end{figure}

%This approach differs from that of a traditional computing architecture, where for all operations, data is transferred from the memory to the \ac{CPU}. 

\ac{NVM} crossbar arrays and analog circuits, however, have inherent non-idealities, such as noise, temporal conductance drift, and non-linear errors, which can lead to imprecision and noisy computation~\cite{BoybatIEDM}. %\textcolor{red}{I would use IBM refs instead, such as JoshiNatComms, BoybatIEDM, TsaiVLSI and Hermes/ARES papers. I also think we should be a bit descriptive of the non-idelities and mention that they also have a temporal evolution}. 
 These effects need to be properly quantified and mitigated to ensure the high accuracy of \ac{DNN} models. In addition to the hardware constraints that are prevalent in edge devices, there is the added complexity of designing \ac{DNN} architectures which are optimized for the edge on a variety of hardware platforms. This requires hardware-software co-design approaches to tackle this complexity, as manually-designed architectures are often tailored for specific hardware platforms. For instance, MobileNet~\cite{Howard2017} uses a depth-wise separable convolution that enhances \ac{CPU} performance but is inefficient for \ac{GPU} parallelization~\cite{Lu2022}. These are bespoke solutions that are often hard to implement and generalize to other platforms.

% The introduction is quite long, and, at least in my opinion, some of this content is more suitable to be included in the related work section.

% \ac{HW-NAS}~\cite{Benmeziane2021} is a promising approach that seeks to automatically identify efficient \ac{DNN} architectures for a target hardware platform. In contrast to traditional \ac{NAS} approaches that focus on searching for the most accurate architectures, \ac{HW-NAS} searches for high-accurate models while optimizing hardware-related metrics such as robustness to noise, latency, power-consumption, and chip area~\cite{Benmeziane2021}.
% Many works~\cite{proxylessnas, fbnet, hwprnas, micronas, micronet} have succeeded in designing \ac{DNN} architectures for tiny and resource-constrained devices using \ac{HW-NAS}. 

% Analog computing brings to bear additional challenges that are not inherent in digital hardware-accelerators such as drift, noise, and device and circuit level variations. 

\ac{HW-NAS}~\cite{Benmeziane2021} is a promising approach that seeks to automatically identify efficient \ac{DNN} architectures for a target hardware platform. In contrast to traditional \ac{NAS} approaches that focus on searching for the most accurate architectures, \ac{HW-NAS} searches for highly accurate models while optimizing hardware-related metrics. Existing \ac{HW-NAS} strategies cannot be readily used with analog \ac{IMC} processors without significant modification for three reasons: (i) their search space contains operations and blocks that are not suitable for analog \ac{IMC}, (ii) lack of a benchmark of hardware-aware trained architectures, and  
% Such operations include depthwise seperable convolutions~\cite{Howard2017} and efficientnet block~\cite{efficientnetv2}, and 
(iii) their search strategy does not include noise injection and temporal drift on weights. % which account for robust and resilient \ac{DNN} architectures on Analog hardware. 

To address these challenges, we propose \textit{AnalogNAS}, a novel \ac{HW-NAS} strategy to design dedicated \ac{DNN} architectures for efficient deployment on edge-based analog \ac{IMC} inference accelerators. This approach considers the inherent characteristics of analog \ac{IMC} hardware in the search space and search strategy.
% AnalogNAS considers the inherent characteristics of Analog hardware to build the search space. 
% An evolution algorithm is used to explore the search space, which simulates biological evolution, including specialized mutations and selection. To compare different sampled networks, we design a surrogate model, able to predict the rank of the architectures, as well as the drift effect after a month of inference. 
% To verify the effectiveness of AnalogNAS, we perform exhaustive hardware simulations, and deploy a subset of network architectures on a real \ac{PCM}-based \ac{IMC} inference accelerator.
Fig.~\ref{fig:motivation} depicts the necessity of our approach. As can be seen, when traditional \ac{DNN} architectures are deployed on analog \ac{IMC} hardware, non-idealities, such as conductance drift, drastically reduce network performance. Networks designed by \textit{AnalogNAS} are extremely
robust to these non-idealities and have much fewer parameters compared to equivalently-robust traditional networks. Consequently, they have reduced resource utilization.

Our specific contributions can be summarized as follows:
\begin{itemize}
    \item We design and construct a search space for analog \ac{IMC}, which contains ResNet-like architectures, including ResNext~\cite{resnext} and Wide-ResNet~\cite{wideresnet}, with blocks of varying widths and depths;
    \item We train a collection of networks using \ac{HWA} training for image classification, \ac{VWW}, and \ac{KWS} tasks. Using these networks, we build a surrogate model to rank the architectures during the search and predict robustness to conductance drift; %Architecture training is performed using \ac{HWA};
    \item We propose a global search strategy that uses evolutionary search to explore the search space and efficiently finds the right architecture under different constraints, including the number of network parameters and analog tiles;
    \item We conduct comprehensive experiments to empirically demonstrate that AnalogNAS can be efficiently utilized to carry out architecture search for various edge tiny applications, and investigate what attributes of networks make them ideal for implementation using analog AI;
    \item We validate a subset of networks on hardware using a 64-core \ac{IMC} chip based on \ac{PCM}.
    % , including image classification, visual wake words, and keyword spotting. By analyzing the search population, we validate the intuition that the search tends to have wide architectures. 
\end{itemize}
% The following paragraph can be removed to save space, if required.
The rest of the paper is structured as follows. In Section~\ref{related_works}, we present related work. In Section~\ref{background}, relevant notations and terminology are introduced. In Section~\ref{analog-nas}, the search space and surrogate model are presented. In Section~\ref{search-strategy}, the search strategy is presented. In Section~\ref{experiment}, the methodology for all experiments is discussed. The simulation results are presented in Section~\ref{results}, along with the hardware validation and performance estimation in Section~\ref{hardware}. The results are discussed in Section~\ref{discussion}. Section~\ref{conclusion} concludes the paper.

\section{Related Work}
\label{related_works}
%In this section, we review related \ac{NAS} work for TinyML and mixed-signal \ac{IMC} %accelerators.

\subsection{NAS for TinyML}
\ac{HW-NAS} has been successfully applied to a variety of edge hardware platforms~\cite{Benmeziane2021, nassurvey}  used to deploy networks for TinyMLPerf tasks~\cite{mlperf} such as image classification, \ac{VWW}, \ac{KWS}, and anomaly detection. %These tasks run properly on resource-constrained devices with less than 10W. 
% \ac{TinyML} allows implementing \ac{DNN} locally on resource-constrained and low-power devices.
% In these hardware, triggering tasks, such as image classification and keyword spotting, are executed. 
% \ac{HW-NAS} is extremely important in this context to find small and efficient \ac{DNN}. TinyMLPerf~\cite{mlperf} defined a benchmark dedicated for \ac{TinyML}. The benchmark contains four tasks: image classification on CIFAR-10, visual wake words, keyword spotting, and anomaly detection. These tasks run properly on resource-constrained devices with less than 10W.
% \newpage
% \paragraph{\ac{NAS} for TinyML} Several works~\cite{micronas, micronet} search for the most efficient \ac{DNN} for microcontrollers and tinyML. 
MicroNets~\cite{micronet} leverages \ac{NAS} for \ac{DL} model deployment on micro-controllers and other embedded systems. It utilizes a differentiable search space~\cite{darts} to find efficient architectures for different TinyMLPerf tasks. For each task, the search space is an extension of current \ac{SOTA} architectures. 
% For instance, they use MobilenetV2~\cite{Howard2017} for Visual Wake Words and modify the width of the first and last convolution in each block to create the search space. While this strategy gave good results, the search space is not flexible as it is based on a single model.
$\mu$-nas~\cite{micronas} includes memory peak usage and a number of other parameters as constraints. Its search strategy combines aging evolution and Bayesian optimization to estimate the objectives and explore a granular search space efficiently. It constructs its search space from a standard \ac{CNN} and modifies the operators' hyper-parameters and a number of layers.

\subsection{NAS for Mixed-Signal IMC Accelerators}
Many works~\cite{flash, nacim, nas4rram, uncertaintyaware} target \ac{IMC} accelerators using \ac{HW-NAS}. FLASH~\cite{flash} uses a small search space inspired by DenseNet~\cite{densenet} and searches for the number of skip connections that efficiently satisfy the trade-off between accuracy, latency, energy consumption, and chip area. Its surrogate model uses linear regression and the number of skip connections to predict model accuracy. NAS4RRAM~\cite{nas4rram} uses \ac{HW-NAS} to find an efficient \ac{DNN} for a specific \ac{RRAM}-based accelerator. It uses an evolutionary algorithm, trains each sampled architecture  without \ac{HWA} training, and evaluates each network on a specific hardware instance.  
% This process allows the search to only find architectures with 88\% accuracy on CIFAR-10. 
NACIM~\cite{nacim} uses co-exploration strategies to find the most efficient architecture and the associated hardware platform. For each sampled architecture, networks are trained considering noise variations. This approach is limited by using a small search space due to the high time complexity of training. UAE~\cite{uncertaintyaware} uses a Monte-Carlo simulation-based experimental flow to measure the device uncertainty induced to a handful of \acp{DNN}. Similar to NACIM~\cite{nacim}, evaluation is performed using \ac{HWA} training with noise injection.
AnalogNet~\cite{analognet} extends the work of Micronet by converting their final models to analog-friendly models, replacing depthwise convolutions with standard convolutions and tuning hyperparameters.
%\textcolor{blue}{We should clearly explain here why AnalogNAS is better (or at least how it is novel), compared to the related work. A sentence or two should be sufficient.}
% Response :

Compared to the above-mentioned \ac{SOTA} HW-NAS strategies, our AnalogNAS is better tailored to analog \ac{IMC} hardware for two reasons: (i) Our search space is much larger and more representative, featuring resnet-like connections. This enables us to answer the key question of what architectural characteristics are suitable for analog \ac{IMC} which cannot be addressed with small search spaces. (ii) We consider the inherent characteristics of analog \ac{IMC} hardware directly in the objectives and constraints of our search strategy in addition to noise injection during the \ac{HWA} training as used by existing approaches.
%\paragraph{Hardware Design Exploration}
%Other works~\cite{} solely focus on designing an efficient chip disregarding the application that will be deployed in it. 

\section{Preliminaries}
\label{background}
\tabulinesep=1mm
\subsection{Analog IMC Accelerator Mechanisms}
\begin{figure}[!b]
    \centering
    \includegraphics[width=0.3\textwidth]{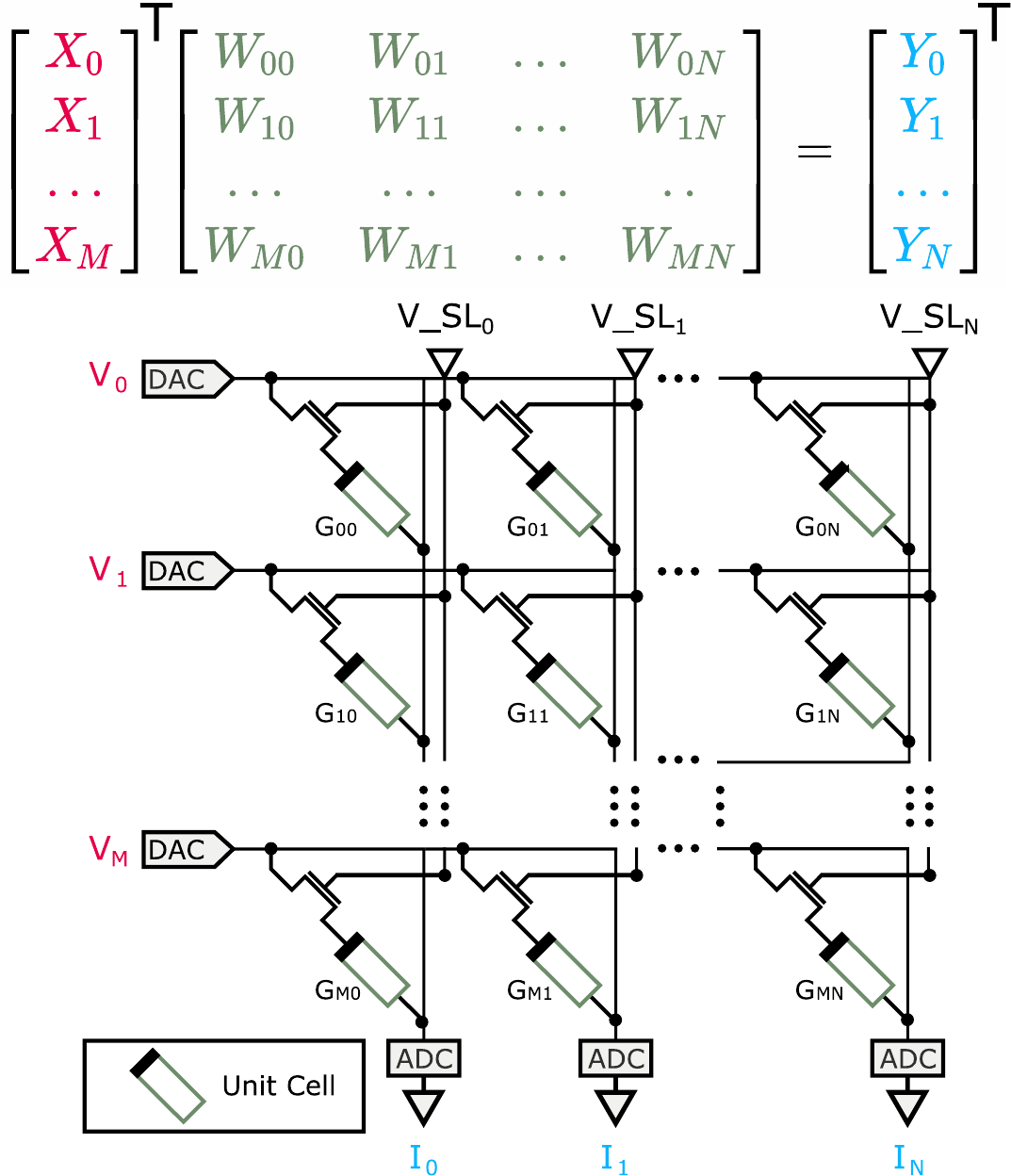}
    \caption{Employed analog \ac{IMC} tile and weight mapping scheme.}
    \label{fig:employed_weight_mapping_scheme}
\end{figure}

Analog \ac{IMC} accelerators are capable of performing \ac{MVM} operations $\mathbf{Y}^T = \mathbf{X}^T\mathbf{W}$ using the laws of physics,  where $\mathbf{W}$ is an $M \times N$ matrix, $\mathbf{X}$ is a  $M \times 1$ vector, and $\mathbf{Y}$ is a $N \times 1$ vector. When arranged in a crossbar configuration, $M \times N$, \ac{NVM} devices can be used to compute \ac{MVM} operations. This is done by encoding elements of $\mathbf{X}$ as \ac{WL} voltages, denoted using $\mathbf{V}$, and elements of $\mathbf{W}$ as conductances of the unit cells, denoted using $\mathbf{G}$.
Negative conductance states cannot be directly encoded/represented using \ac{NVM} devices. Consequently, differential weight mapping schemes are commonly employed, where either positive weights, i.e., $\mathbf{W^{+}} = \max(\mathbf{W}, 0)$, and negative weights, i.e., $\mathbf{W^{-}} =  -\min(\mathbf{W}, 0)$, are encoded within unit cells, using alternate columns, or on different tiles~\cite{Lammie2022}.
% A crossbar array contains $\#\,\text{Row} \times \#\,\text{Col}$ cells and the peripheral devices ... ???? .... The cell can be of different types: PCM, RRAM, and MRAM. The weights ($W$) in the matrix are mapped to the conductance of the cells $G$ ... What is G ?. 
% Since the computation is in analog, the input vector $x$ is transferred to analog voltage $V$ using the \ac{DAC}. 
The analog computation, i.e., $\mathbf{I}=\mathbf{VG}$ is performed, where the current flow to the end of the $N$-th column is $I_N = \sum_{i=0}^MG_{i,N}V_i$. Typically, \acp{DAC} are required to encode \ac{WL} voltages and \acp{ADC} are required to read the output currents of each column. The employed analog \ac{IMC} tile, its weight mapping scheme, and computation mechanism are depicted in Fig.~\ref{fig:employed_weight_mapping_scheme}.

\subsection{Temporal Drift of Non-Volatile Memory Devices}\label{temporal_evolution}
Many types of \ac{NVM} devices, most prominantly, \ac{PCM}, exhibit temporal evolution of the conductance values referred to as the conductance drift. This poses challenges for maintaining synaptic weights reliably~\cite{DBLP:journals/corr/abs-1906-03138}. Conductance drift is most commonly modelled using Eq.~(\ref{drift}), as follows:
\begin{equation}\label{drift}
    G(t) = G(t_0)(t/t_0)^{-\nu},
\end{equation}
where $G(t_0)$ is the conductance at time $t_0$ and $\nu$ is the drift exponent. In practice, conductance drift is highly stochastic because $\nu$ depends on the programmed conductance state and varies across devices. Consequently, when reporting the network accuracy at a given time instance (after device programming), it is computed across multiple experiment instances (trials) to properly capture the amount of accuracy variations.

\subsection{HWA-training and analog hardware accuracy evaluation simulation}
To simulate training and inference on analog \ac{IMC} accelerators, the \ac{AIHWKIT}~\cite{aihwkit} is used. The \ac{AIHWKIT} is an open-source Python toolkit for exploring and using the capabilities of in-memory computing devices in the context of artificial intelligence and has been used for \ac{HWA} training of standard \acp{DNN} with hardware-calibrated device noise and drift models~\cite{aihwkit_arxiv}.
% extend 

\subsection{Hardware-aware Neural Architecture Search (HW-NAS)}
\ac{HW-NAS} refers to the task of automatically finding the most efficient \ac{DNN} for a specific dataset and target hardware platform.
\ac{HW-NAS} approaches often employ black-box optimization methods such as evolutionary algorithms~\cite{eeea}, reinforcement learning~\cite{fbnet,DBLP:conf/dac/JiangZSYZSH19}, and Bayesian optimization~\cite{bayesianhwnas, hao}. The optimization problem is either cast as a constrained or multi-objective optimization~\cite{Benmeziane2021}. 
In AnalogNAS, we chose constrained optimization over multi-objective optimization for several reasons. First, constrained optimization is more computationally efficient than multi-objective optimization, which is important in the context of \ac{HW-NAS}, to allow searching a large search space in a practical time frames. %\textcolor{red}{Why is it more important for HW-NAS to be computationally efficient?}. 
Multi-objective optimization is computationally expensive and can result in a large number of non-dominated solutions that can be difficult to interpret. Secondly, by using constrained optimization, we can explicitly incorporate the specific constraints of the analog hardware in our search strategy. This enables us to find \ac{DNN} architectures that are optimized for the unique requirements and characteristics of analog \ac{IMC} hardware, rather than simply optimizing for multiple objectives.

%to a single objective function using scalarization, such as a weighted sum of the objectives, i.e., task-specific performance and hardware efficiency. The weights are usually fixed via empirical tests. In this case, the result is a single architecture that maximizes the objective. 

%\begin{equation}
%    \min_{\alpha \in A} f_1(\alpha),\dots,f_n(\alpha),
%    \label{eq:bg_1}
%\end{equation}

%\noindent where $A$ referrers to the architecture search space, $\alpha$ to one sampled architecture, and $f_i$ is the function that gives performance metric $i$, where $i$ may represent accuracy, latency, energy consumption, or memory occupancy, etc.

%During the search process, each architecture undergoes an evaluation of its objective functions. However, as training a single architecture could span over several hours, the convergence of Hardware Neural Architecture Search (HW-NAS) can take a considerable amount of time. To tackle this issue, evaluation methods have evolved into estimation strategies utilizing surrogate models. We built a machine learning predictor per objective to estimate the analog-aware accuracy and noise robustness. 

\section{Analog-NAS}
\label{analog-nas}
The objective of AnalogNAS is to find an efficient network architecture under different analog \ac{IMC} hardware constraints. AnalogNAS comprises three main components: (i) a resnet-like search space, (ii) an analog-accuracy surrogate model, and (iii) an evolutionary-based search strategy. We detail each component in the following subsections.

\subsection{Resnet-like Search Space}
Resnet-like architectures have inspired many manually designed \ac{SOTA} \ac{DL} architectures, including Wide ResNet~\cite{wideresnet} and EfficientNet~\cite{efficientnetv2}. Their block-wise architecture offers a flexible and searchable macro-architecture for \ac{NAS}~\cite{nasnet}. Resnet-like architectures can be implemented efficiently using \ac{IMC} processors, as they are comprised of a large number of \ac{MVM} and element-wise operations. Additionally, due to the highly parallel nature of IMC, Resnet architectures can get free processing of additional input/output channels. This makes Resnet-like architectures highly amenable to analog implementation.%\textcolor{red}{should we say they don't have DWconv, so analogamenable fully?}.

\begin{figure}[!t]
    \centering%https://docs.google.com/presentation/d/1KB9FJjv66OzRH1QKikQPgVkVS9dnhkMqYhtXzDVNEKY/edit?usp=sharing
    \includegraphics[width=0.4\textwidth]{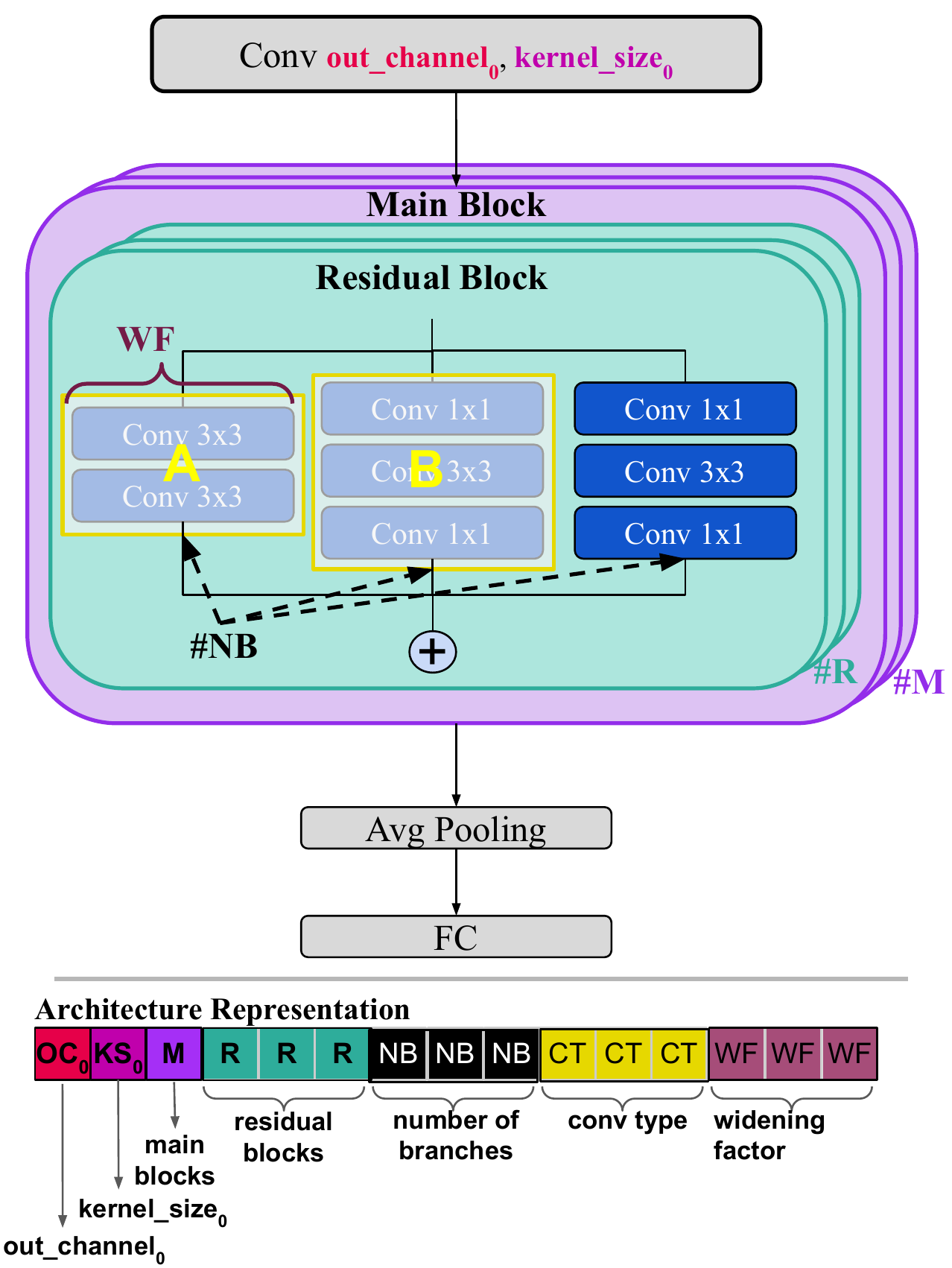}
    \caption{Resnet-like macro architecture. %\textcolor{blue}{Are the "A" and "B" labels in white intentional?. If so, these are confusing and should either be relocated/explained or removed.}
    }
    \label{fig:macro_architecture}
\end{figure}

%Our search space is inspired by the Resnet architecture. 
Fig.~\ref{fig:macro_architecture} depicts the macro-architecture used to construct all architectures in our search space. %The searchable elements are illustrated in blue. 
The architecture consists of a series of $M$ distinct main blocks. %\textcolor{red}{does this imply that we can only repeat the same M block? I assume not, from the apction of Table II} 
Each main block contains $R$ residual blocks. The residual blocks use skip connections with or without downsampling. Downsampling is performed using 1x1 convolution layers when required, i.e., when the input size does not match the output size. The residual block can have $B$ branches. Each branch uses a convolution block. We used different types of convolution blocks to allow the search space to contain all standard architectures such as Resnets~\cite{resnet}, ResNext~\cite{resnext}, and Wide Resnets~\cite{wideresnet}. The standard convolution blocks used in Resnets, commonly referred to as \textit{BottleNeckBlock} and \textit{BasicBlock}, are denoted as A and B respectively. We include variants of A and B in which we inverse the order of the ReLU and Batch normalization operations. The resulting blocks are denoted as C and D. Table~\ref{tab:search_space} summarizes the searchable hyper-parameters and their respective ranges.
The widening factor scales the width of the residual block. We sample architectures with different depths by changing the number of main and residual blocks. The total size of the search space is approximately 73B architectures. The larger architecture would contain 240 convolutions and start from an output channel of 128 multiplying that by 4 for every 16 blocks.

\subsection{Analog-accuracy Surrogate Model}\label{sec:surrogate} 
\subsubsection{Evaluation Criteria}
To efficiently explore the search space, a search strategy requires evaluating the objectives of each sampled architecture. 
Training the sampled architectures is very time-consuming; especially when \ac{HWA} retraining is performed, as noise injection and I/O quantization modeling greatly increases the computational complexity.
Consequently, we build a surrogate model capable of estimating the objectives of each sampled architecture in \ac{IMC} devices. 
To find architectures that maximize accuracy, stability, and resilience against \ac{IMC} noise and drift characteristics, we have identified the following three objectives.

\paragraph{The 1-day accuracy} is the primary objective that most \ac{NAS} algorithms aim to maximize. It measures the performance of an architecture on a given dataset. When weights are encoded using \ac{IMC} devices, the accuracy of the architecture can drop over time due to conductance drift. Therefore, we have selected the 1-day accuracy as a metric to measure the architecture's performance. %By focusing on the 1-day accuracy, we can evaluate the performance of the architecture over a shorter period and mitigate the effects of conductance drift. 

\paragraph{The \acf{AVM}} is the difference between the 1-month and 1-sec accuracy. This objective is essential to measure the robustness over a fixed time duration. A 30-day period allows for a reasonable trade-off between capturing meaningful accuracy changes and avoiding short-term noise and fluctuations that may not reflect long-term trends.

\paragraph{The 1-day accuracy standard deviation} measures the variation of the architecture's performance across experiments, as discussed in Section~\ref{temporal_evolution}. A lower standard deviation indicates that the architecture produces consistent results on hardware deployments, which is essential for real-world applications.

\noindent To build the surrogate model, we follow two steps: Dataset creation and Model training:
\subsubsection{Dataset Creation}
The surrogate model will predict the rank based on the 1-day accuracy and estimates the \ac{AVM} and 1-day accuracy standard deviation using the \ac{MSE}. Since the search space is large, care has to be taken when sampling the dataset of architectures that will be used to train the surrogate model. 

\begin{table}[!t]
\caption{Searchable hyper-parameters and their respective ranges.}
\begin{threeparttable}
\begin{tabular}{p{0.05\textwidth}p{0.18\textwidth}p{0.18\textwidth}}
\toprule \toprule
\textbf{Hyper-parameter} & \textbf{Definition}& \textbf{Range}                            \\ \midrule
OC\textsubscript{0}    & First layer's output channel      & Discrete Uniform [8, 128] \\ %\midrule
KS\textsubscript{0}   & First layer's kernel size        & Discrete Uniform [3, 7]   \\ %\midrule
M     & Number of main blocks                  & Discrete Uniform [1, 5]  \\ %\midrule
R*        & Number of residual block per main block              & Discrete Uniform [1, 16] \\ %\midrule
B*          & Number of branches per main block           & Discrete Uniform [1, 12]  \\ %\midrule
CT*    &    Convolution block type per main block       & Uniform Choice [A; B; C; D]                 \\ %\midrule
WF*   &   Widening factor per main block  & Uniform [1, 4]            \\ \bottomrule \bottomrule
\end{tabular}
\begin{tablenotes}
\setlength\labelsep{0pt}
\medskip
    \item *The hyper-parameter is repeated for each main block. ConvBlock refers to different Conv-Relu-BN blocks. 
\end{tablenotes}
\end{threeparttable}
\label{tab:search_space}
\end{table}

\begin{figure}[!b]
    \centering
    \includegraphics[width=0.4\textwidth]{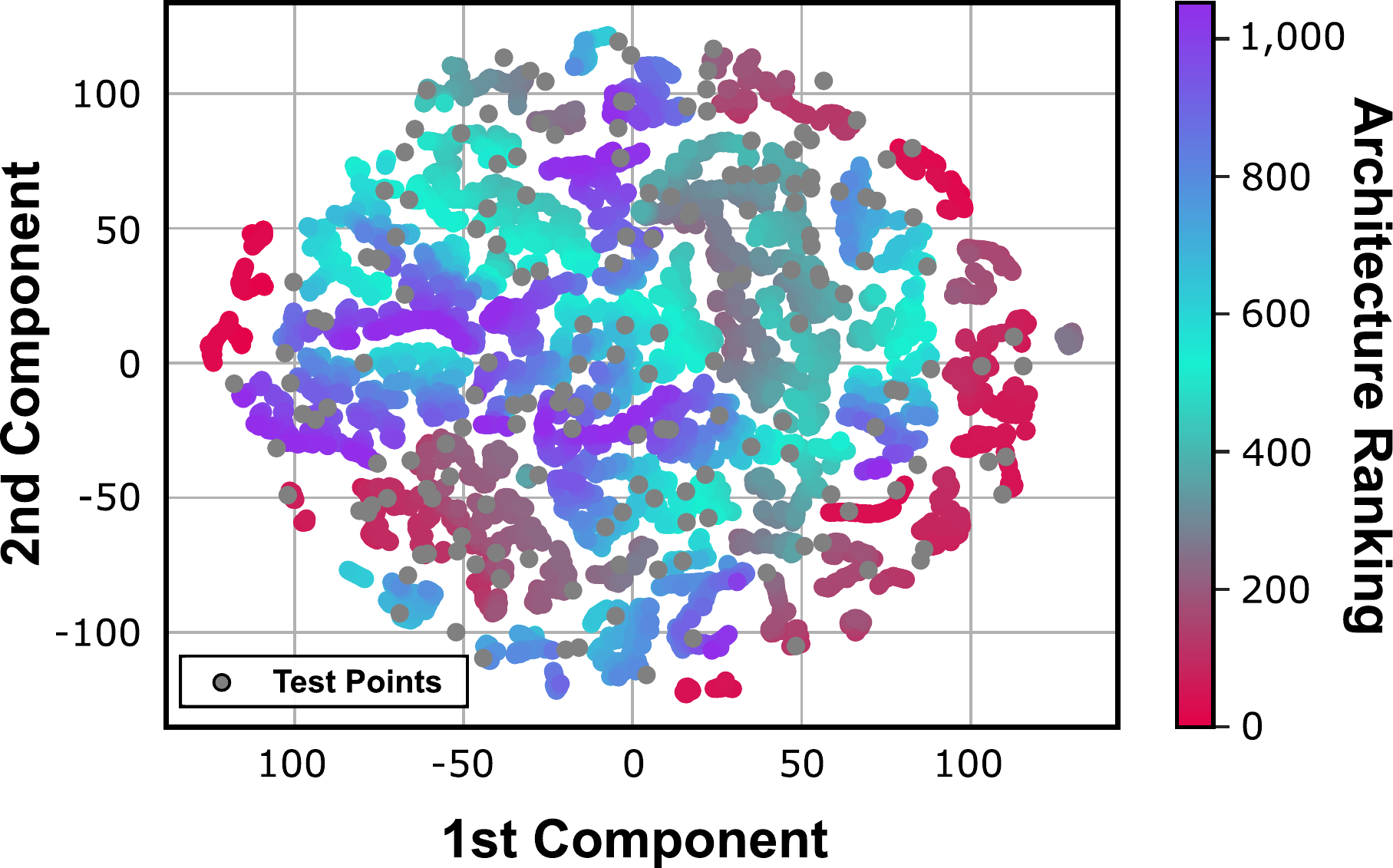}
    \caption{t-Distributed Stochastic Neighbor Embedding (t-SNE) visualization of the sampled architectures for CIFAR-10. %\textcolor{blue}{Please update this figure.}
    }
    \label{fig:tsne}
\end{figure}

%\subsection{Sampling}
The architectures of the search space are sampled using two methods: (i) \ac{LHS}~\cite{lhs} and (ii) \ac{NAS} with full training. A more detailed description of the AnalogNAS algorithm is presented in Section~\ref{search-strategy}.
We use \ac{LHS} to sample architectures distributed evenly over the search space. This ensures good overall coverage of different architectures and their accuracies. \ac{NAS} with full training is performed using an evolutionary algorithm to collect high-performance architectures. This ensures good exploitation when reaching well-performing regions.
% \begin{itemize}
%     \item \textit{Latin Hypercube Sampling~\cite{} (LHS)}. We use LHS to sample architectures distributed evenly over the search space. This ensures having good overall coverage of different architectures and their accuracies.
%     \item \textit{Neural Architecture Search with full training.} We use an evolutionary algorithm to collect high-performance architectures. This ensures good exploitation when reaching well-performing regions.  
% \end{itemize}
In Fig.~\ref{fig:tsne}, we present a visualization of the search space coverage, which does not show any clustering of similarly performing architectures at the edge of the main cloud of points. Thus, it is not evident that architectures with similar performance are located close to each other in the search space.  This suggests that traditional search methods that rely on local optimization may not be effective in finding the best-performing architectures. Instead, population-based search strategies, which explore a diverse set of architectures, could be more effective in finding better-performing architectures. Our search strategy extracted 400 test points, and we found that architectures were distributed throughout the main cloud, indicating that our dataset covers a diverse portion of the search space, despite the limited size of only 1,000. %\textcolor{red}{Do we need to give any details on the LHS and NAS parameters/methods?}

%\subsection{Training}
Each sampled architecture is trained using different levels of weight noise and \ac{HWA} training hyper-parameters using the \ac{AIHWKIT}~\cite{aihwkit}. Specifically, we modify the standard deviation of the added weight noise between [0.1, 5.0] in increments of 0.1. The tile size was assumed to be symmetric and varied in [256, 512], representing 256-by-256 and 512-by-512 arrays respectively. %In addition, the weight noise factor was modified between [0.1, 0.5] \textcolor{red}{Didn't we mention this two sentences ago?}. 
Training with different configurations allowed us to generalize the use of the surrogate model across a range of IMC hardware configurations, and to increase the size of the constructed dataset.

\subsubsection{Model training}
To train the surrogate model, we used a hinge pair-wise ranking loss~\cite{gates} with margin $m=0.1$. The hinge loss, defined in Eq. (\ref{eq:loss}), allows the model to learn the relative ranking order of architectures rather than the absolute accuracy values~\cite{gates,hwprnas}.

\begin{equation}
    L(\{a_j, y_j\}_{j=1,...,N}) = \sum^{N}_{j=1} \sum_{\{i,j| y_i > y_j\}} \max[0, m - P(a_i) - P(a_j)]
    \label{eq:loss}
\end{equation}

\begin{figure}[!t]
    \centering
    \includegraphics[width=0.43\textwidth]{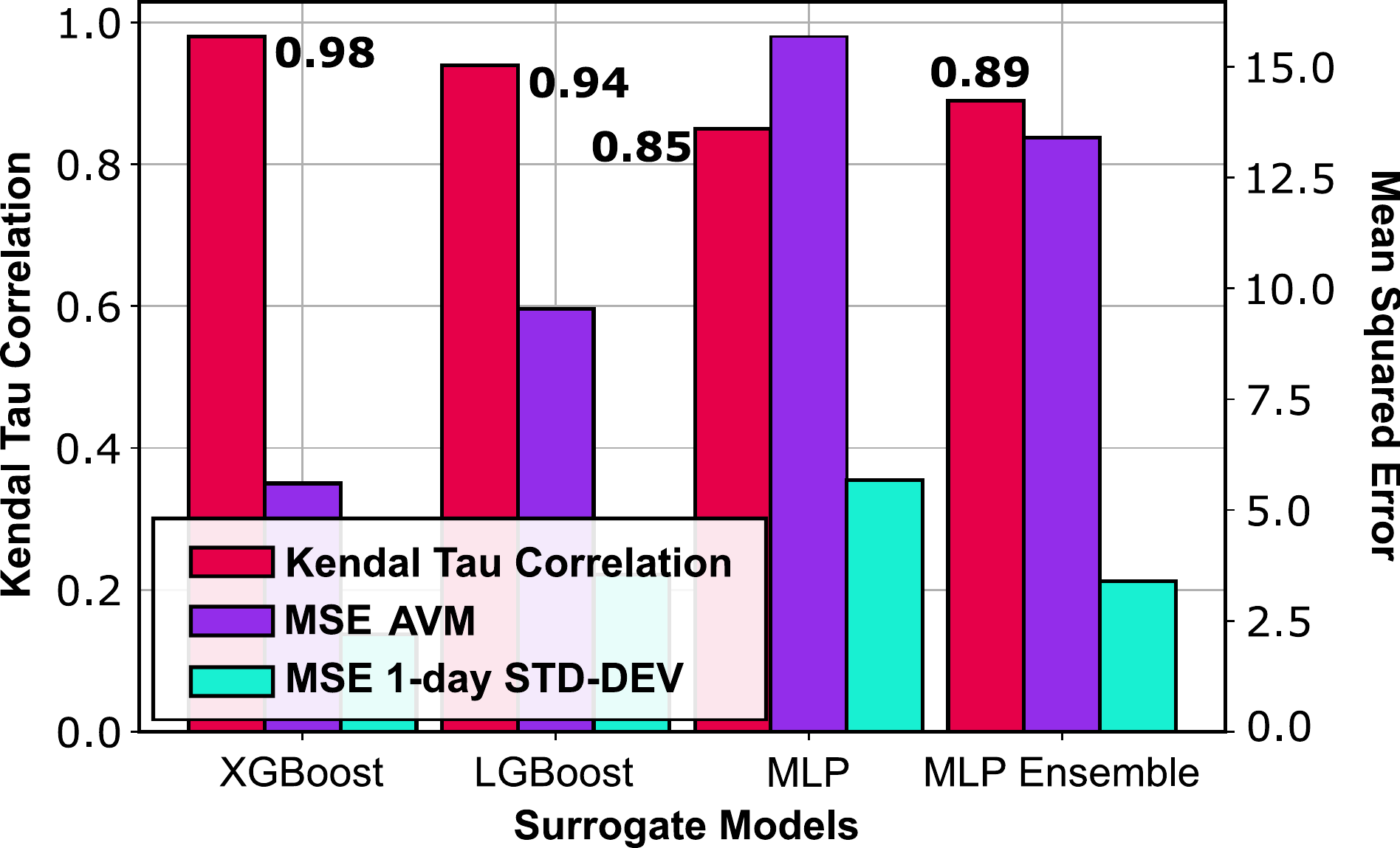}
    \caption{Surrogate models comparison.}
    \label{fig:surrogate_models}
    \vspace{-0.5cm}
\end{figure}

\noindent $a_j$ refers to architectures indexed $j$, and $y_j$ to its corresponding 1-day accuracy. $P(a)$ is the predicted score of architecture $a$. $P(a)$ during training, the output score is trained to be correlated with the actual ranks of the architectures. 
Several algorithms were tested. After an empirical comparison, we adopted Kendall's Tau ranking correlation~\cite{kendall} as the direct criterion for evaluating ranking surrogate model performance. Fig.~\ref{fig:surrogate_models} shows the comparison using different \ac{ML} algorithms to predict the rankings and \acp{AVM}. Our dataset is tabular. It contains each architecture and its corresponding features. XGBoost outperforms the different surrogate models in predicting the architectures' ranking order, the \ac{AVM} of each architecture, and the 1-day standard deviation. 

\section{Search Strategy}\label{search-strategy}
\begin{figure}
    \centering
    %https://docs.google.com/presentation/d/1f3UkRIYihthNVclDd1GpaedY6dcwV7nDW2wzQWCWQ84/edit?usp=sharing
    \includegraphics[width=0.45\textwidth]{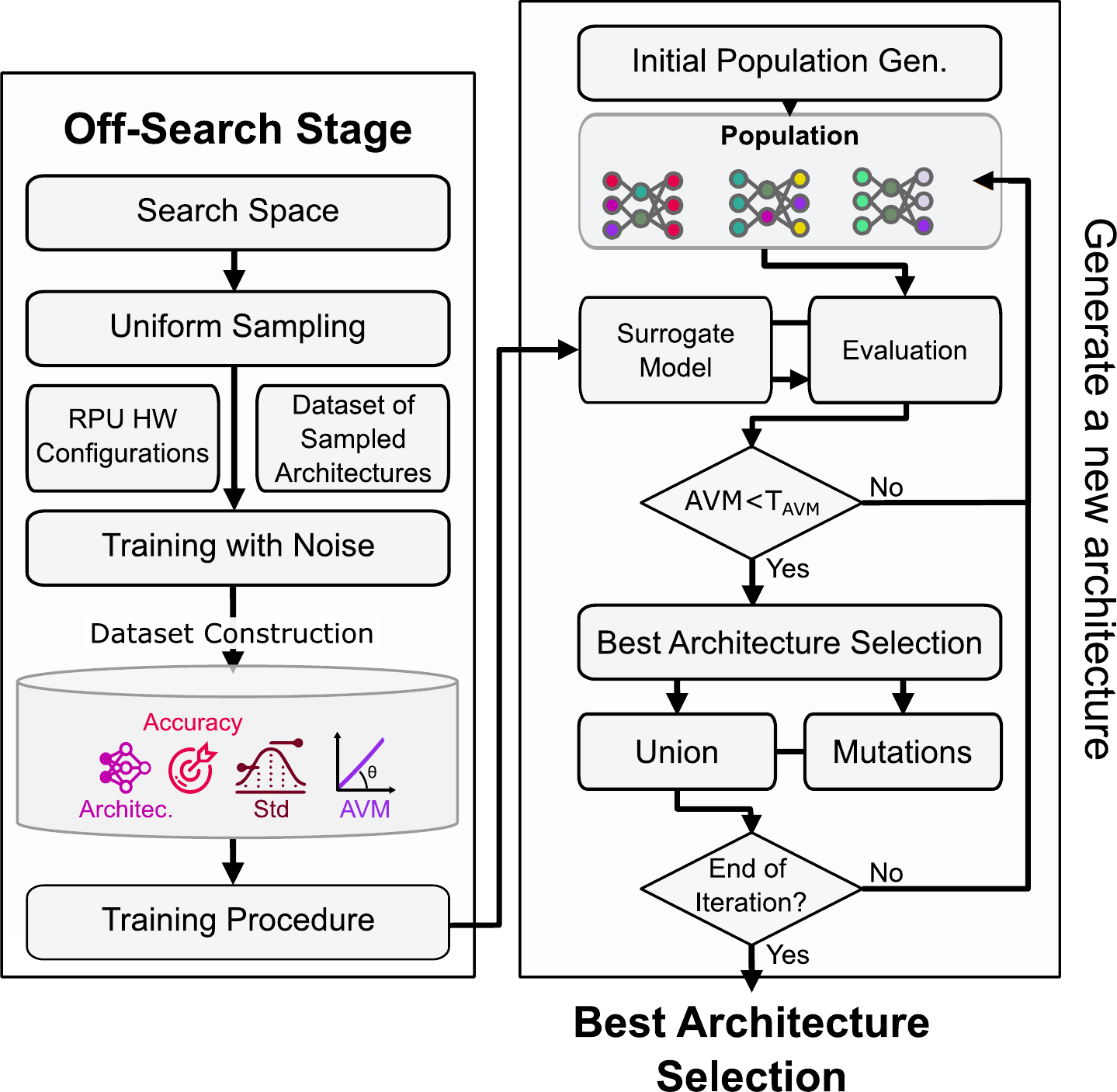}
    \caption{Overview of the AnalogNAS framework.}
    \label{fig:analog_nas}
\end{figure}

Fig.~\ref{fig:analog_nas} depicts the overall search framework. 
Given a dataset and a hardware configuration readable by \ac{AIHWKIT}, the framework starts by building the surrogate model presented in Section~\ref{sec:surrogate}. Then, we use an optimized evolutionary search to efficiently explore the search space using the surrogate model.
Similar to traditional evolutionary algorithms, we use real number encoding. Each architecture is encoded into a vector, and each element of the vector contains the value of the hyper-parameter, as listed in Table~\ref{tab:search_space}. 
% Our evolutionary search algorithm is formally defined using Algorithm~\ref{alg:ea_algorithm}.

\subsection{Problem Formulation}
%Our search strategy is generalized on four TinyMLPerf tasks: (i) image classification, (ii) \ac{VWW}, (iii) \ac{KWS}, and (iv) anomaly detection.
% MicroNet~\cite{} uses \ac{HW-NAS} to find an efficient \ac{DNN} architecture for each of these tasks for micro-controllers. 
% In this context, the architectures are small enough to be deployed in tiny devices. Therefore,
Given the search space $S$, our goal is to find an architecture $\alpha$, that maximizes the 1-day accuracy while minimizing the 1-day standard deviation, subject to constraints on the number of parameters and the \ac{AVM}. The number of parameters is an important metric in \ac{IMC}, because it directly impacts the amount of on-chip memory required to store the weights of a \ac{DNN}. Eq.~(\ref{eq1}) formally describes the optimization problem as follows: %\textcolor{blue}{Please check that this equation is formatted correctly.}
\begin{equation} \label{eq1}
\begin{split}
\max_{\alpha \in S }\; &  \qquad\frac{\textnormal{ACC}(\alpha)}{\sigma(\alpha)} \\
 \textnormal{s.t} & \qquad\psi(\alpha) < T_p\text{,}\qquad\textnormal{AVM}(\alpha) < T_{\text{AVM}} \\
\end{split}
\end{equation}

 ACC refers to the 1-day accuracy objective, $\sigma$ denotes the 1-day accuracy's standard deviation, and $\psi$ is the number of parameters. $T_p$ and $T_{\text{AVM}}$ are user-defined thresholds that correspond to the maximum number of parameters and \ac{AVM}, respectively.

\subsection{Search Algorithm}
Our evolutionary search algorithm, i.e., AnalogNAS, is formally defined using Algorithm~\ref{alg:ea_algorithm}.
AnalogNAS is an algorithm to find the most accurate and robust neural network architecture for a given analog \ac{IMC} configuration and task. % \textcolor{red}{why HW device? Device technology/device perhaps? We do not have any constraint on HW perf/power metrics here in the search.}. 
The algorithm begins by generating a dataset of neural network architectures, which are trained on the task and evaluated using \ac{AIHWKIT}. %\textcolor{red}{We do not do any perf/power efficiency evaluations? Perhaps accuracy? Comment also applies to the next sentence}.
A surrogate model is then created to predict the efficiency of new architectures. The algorithm then generates a population of architectures using an \ac{LHS} technique and selects the top-performing architectures to be mutated and generate a new population. The process is repeated until a stopping criterion is met, such as a maximum number of iterations or a time budget. Finally, the most robust %\textcolor{red}{efficient} 
architecture is returned. In the following, we detail how the population initialization, fitness evaluation, and mutations are achieved.

\begin{algorithm}[!t]
   \caption{AnalogNAS algorithm. }
   \label{alg:ea_algorithm}
\begin{algorithmic}
   \STATE {\bfseries Input:} Search space: $S$, RPU Configuration: $rpu\_config$, target task: $task$, population size:  $population\_size$, \ac{AVM} threshold: $T_{\text{AVM}}$, parameter threshold: $T_p$, number of iterations: $N$, $time\_budget$
   \STATE {\bfseries Output:} Most efficient architecture for $rpu\_config$ in $S$

   \STATE \textbf{Begin}
   \STATE $D$ = sample($S$, dataset\_size)
   \STATE HW\_Train($D$, $task$)
   \STATE \acp{AVM} = compute\_AVM($D$)
   \STATE surrogate\_model = XGBoost
   \STATE train(surrogate\_model, $D$, \acp{AVM}) 
   \REPEAT
    \STATE population = LHS(population\_size, $T_p$) 
    \STATE \acp{AVM}, ranks = surrogate\_model(population)
   \UNTIL{\acp{AVM} $> T_{\text{AVM}}$ }
   \WHILE{$i<N$ or $time < time\_budget$ }
   \STATE top\_50 = select(population, ranks) 
   \STATE mutated = mutation(top\_50, $T_p$)
   \STATE population = top\_50 $\bigcup$ mutated 
   \STATE \acp{AVM}, ranks = surrogate\_model(population)
   \ENDWHILE
   \STATE \textbf{return} $top_1$(population, ranks)
\end{algorithmic}
\end{algorithm}

\subsubsection{Population Initialization}
The search starts by generating an initial population. Using the \ac{LHS} algorithm, we sample the population uniformly from the search space. \ac{LHS} ensures that the initial population contains architectures with different architectural features. LHS is made faster with parallelization by dividing the sampling into multiple independent subsets, which can be generated in parallel using multiple threads.

\subsubsection{Fitness Evaluation}
We evaluate the population using the aforementioned analog-accuracy surrogate model. In addition to the rankings, the surrogate model predicts the \ac{AVM} of each architecture. As previously described, the \ac{AVM} is used to gauge the robustness of a given network. If the \ac{AVM} is below a defined threshold, $T_{\text{AVM}}$, the architecture is replaced by a randomly sampled architecture. The new architecture is constrained to be sampled from the same hypercube dimension as the previous one. This ensures efficient exploration.

\subsubsection{Selection and Mutation}
We select the top 50\% architectures from the population using the predicted rankings. These architectures are mutated. The mutation functions are classified as follows:% \textcolor{red}{Can we provide the associated parameters from Fig3 to the items below, e.g., M, R, B, ...)}: 

\paragraph{Depth-related mutations} modify the depth of the architectures. Mutations include adding a main block, by increasing or decreasing $M$ or a residual block $R$, or modifying the type of convolution block, i.e.,  $\{A, B, C, D\}$, for each main block. 
    
\paragraph{Width-related mutations} modify the width of the architectures. Mutations include modifying the widening factor $W$ of a main block or adding or removing a branch $B$, or modifying the initial output channel size of the first convolution, $OC$. 
    
\paragraph{Other mutations} modify the kernel size of the first convolution, $KS$, and/or add skip connections,denoted using $ST$. 

Depth- and width-related mutations are applied with the same probability of 80\%. % \textcolor{red}{What is the probability?}. 
The other mutations are applied with a 50\% probability. %\textcolor{red}{is this in terms of percentage? or is it 50 percent?}. 
In each class, the same probability is given to each mutation. The top 50\% architectures in addition to the mutated architectures constitute the new population.
For the remaining iterations, we verify the ranking correlation of the surrogate model. If the surrogate model's ranking correlation is degraded, we fine-tune the surrogate model with the population's architectures. The degradation is computed every 100 iterations. The surrogate model is tested on the population architectures after training them. It is fine-tuned if Kendall's tau correlation drops below 0.9. %\textcolor{red}{How do we assess the degradation? How many models from the population are used for re-training? Could we be a bit more quantitative here?}.

%\newpage
%\color{magenta}

\section{Experiments}
\label{experiment}
This section describes the experiments used to evaluate AnalogNAS on three tasks: CIFAR-10 image classification, \ac{VWW}, and \ac{KWS}. The \ac{AIHWKIT} was used to perform hardware simulations. %In addition, the hardware performance of each network was estimated using a proprietary tool entitled \texttt{ADS}~\cite{}. Finally, hardware validation was performed for a small subset of image classification networks using a \ac{PCM}-based \ac{IMC} processor.

\subsection{Experimental Setup}

\subsubsection{Training Details} We detail the hyper-parameters used to train the surrogate model and different architectures on CIFAR-10, \ac{VWW}, and \ac{KWS} tasks.

\paragraph{Surrogate model training} We trained a surrogate model and dataset of \ac{HWA} trained \ac{DNN} architctures for each task. The sizes of the datasets were 1,200, 600, and 1,500, respectively. An additional 500 architectures were collected during the search trials for validation. All architectures were first trained without noise injection (i.e., using vanilla training routines), and then converted to \ac{AIHWKIT} models for \ac{HWA} retraining. The surrogate model architecture used was XGBoost. For \ac{VWW} and \ac{KWS}, the surrogate model was fine-tuned from the image classification XGBoost model.

\paragraph{Image classification training} We first trained the network architectures using the CIFAR-10 dataset~\cite{cifar10}, which contains 50,000 training and 10,000 test samples, evenly distributed across 10 classes. We augmented the training images with random crops and cutouts only. For training, we used \ac{SGD} with a learning rate of 0.05 and a momentum of 0.9 with a weight decay of 5e-4. The learning rate was adjusted using a cosine annealing learning rate scheduler with a starting value of 0.05 and a maximum number of 400 iterations. 

\paragraph{\acf{VWW} training}We first trained the network architectures using the \ac{VWW} dataset~\cite{vww}, which contains 82,783 train and 40,504 test images. Images are labeled 1 when a person is detected, and 0 when no person is present. The image pre-processing pipeline includeded horizontal and vertical flipping, scale augmentation~\cite{resnest}, and random \ac{RGB} color shift. To train the architectures, we used the RMSProp optimizer~\cite{rmsprop} with a momentum of 0.9, a learning rate of 0.01, a batch normalization momentum of 0.99, and a $l_2$ weight decay of 1e-5. 

\paragraph{\acf{KWS} training} We first trained the network architectures using the \ac{KWS} dataset~\cite{speechcommandsv2}, which contains 1-second long incoming audio clips. These are classified into one of twelve keyword classes, including "silence" and "unknown" keywords. The dataset contains 85,511 training, 10,102 validation, and 4,890 test samples. The input was transformed to 49 × 10 × 1 features from the Mel-frequency cepstral coefficients~\cite{melfreq}.
The data pre-processing pipeline included applying background noise and random timing jitter. To train the architectures, we used the Adam optimizer~\cite{adam} with a decay of 0.9, a learning rate of 3e-05, and a linear learning rate scheduler with a warm-up ratio of 0.1. 

\subsubsection{Search Algorithm}
The search algorithm was run five times to compute the variance. The evolutionary search was executed with a population size of 200. If not explicitly mentioned, the \ac{AVM} threshold was set to 10\%. The width and depth mutation probability was set to 0.8. The other mutations' probability was set to 0.5. The total number of iterations was 200. After the search, the obtained architecture for each task was compared to \ac{SOTA} baselines for comparison.
\begin{figure*}[ht!]
    \centering
    \includegraphics[width=0.9\textwidth, clip, trim=0 0 0 0 ]{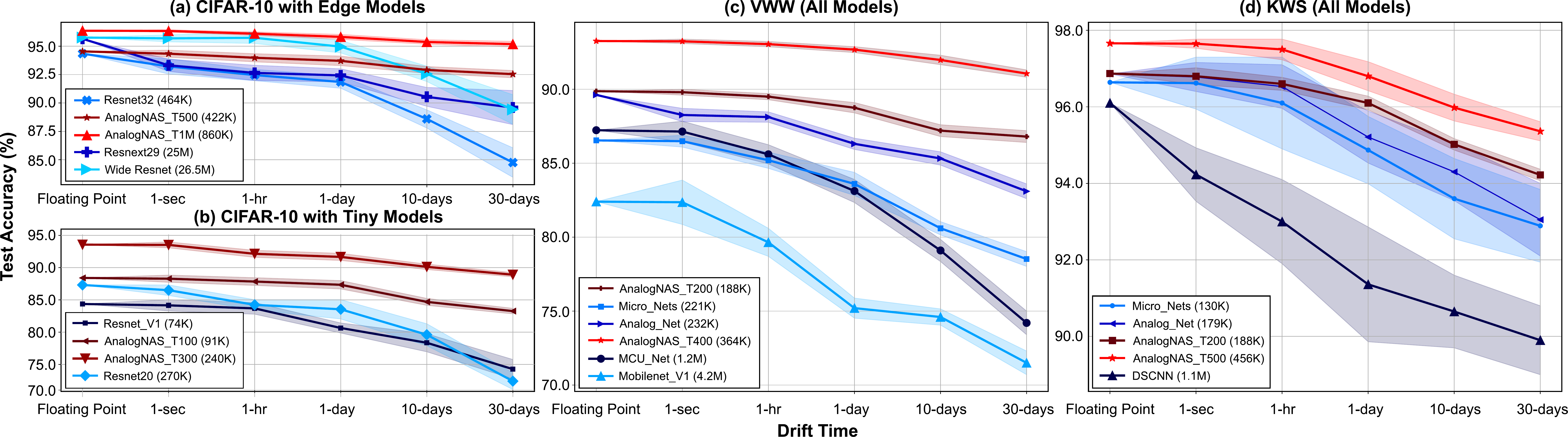}
    \vspace{-0.45cm}
    \caption{Simulated hardware comparison results on three benchmarks: (a,b) CIFAR-10, (c)\ac{VWW}, and (d) \ac{KWS}. The size of the marker represents the size (i.e., the number of parameters) of each model. The shaded area corresponds to the standard deviation at that time.}
    \label{fig:results}
\end{figure*}

\subsection{Results}\label{results}
\begin{figure}[!t]
    \centering
    \includegraphics[width=0.45\textwidth]{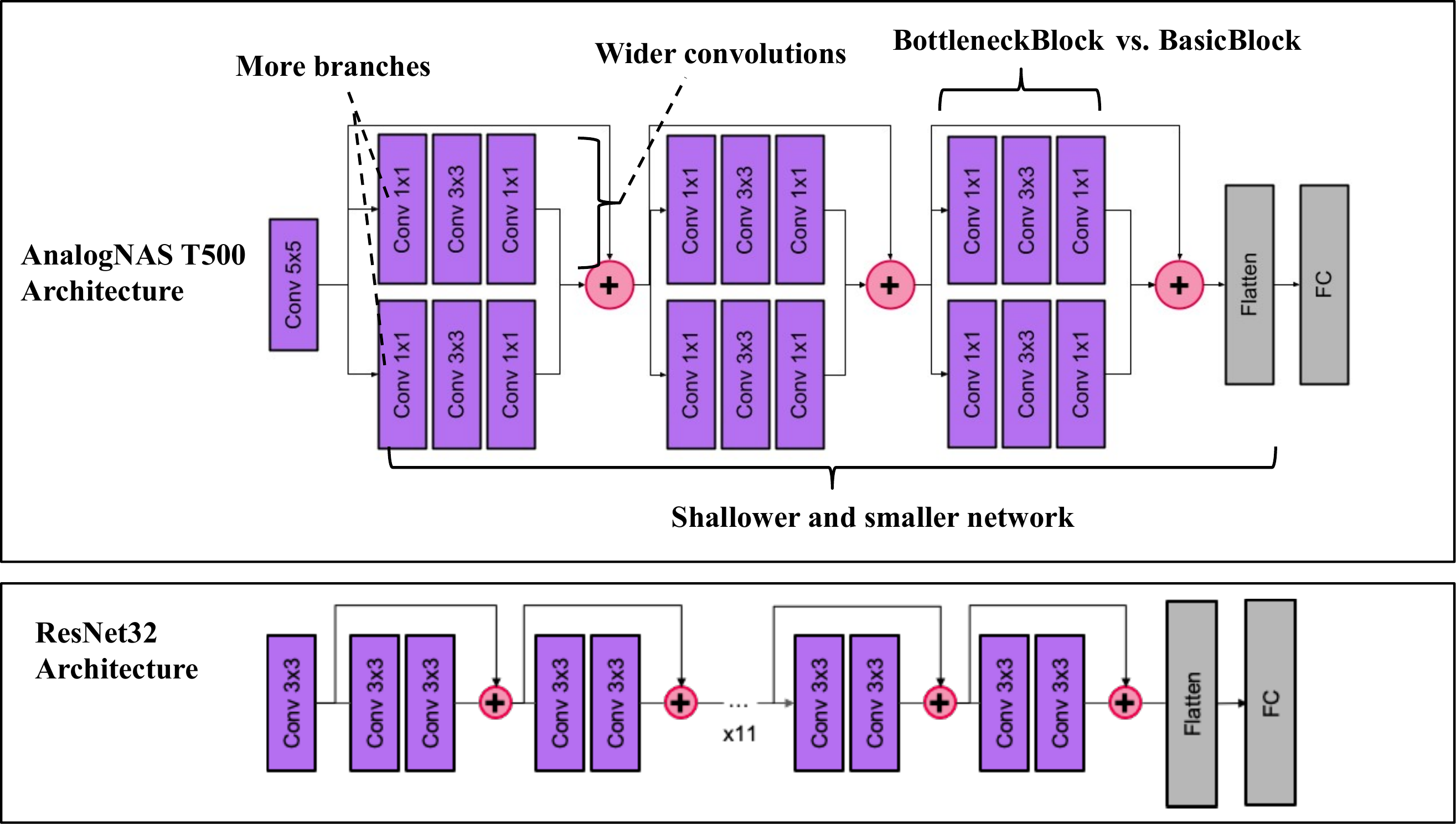}
     \caption{Architectural differences between AnalogNAS\_T500 and Resnet32.}\label{fig:architectures}
\end{figure}

The final architecture compositions for the three tasks are listed in Table~\ref{table:final_networks}. In addition, figure~\ref{fig:architectures} highlights the architectural differences between AnalogNAS\_T500 and resnet32.  
We modified $T_{p}$ to find smaller architectures. To determine the optimal architecture for different parameter thresholds, we use T$X$, where $X$ represents the threshold $T_p$ in K units (e.g., T100 refers to the architecture with a threshold of 100K parameters). When searching for T200 and T100, the probability of increasing the widening factor or depth to their highest values, was lessened to 0.2. %\textcolor{red}{The last sentence is a bit unclear. To what values are they lessened? Also, can we bring the footnote to the next, but in a shorter and more clear way?}.
%  \begin{figure}
%      \centering
%      \includegraphics[clip, trim=0 1cm 0 0, width=0.45\textwidth]{figures/architectures.pdf}
%      \caption{\textcolor{red}{No clue what all those colors, numbers and letters mean. Please define that somewhere, otherwise that figure is impossible to understand and therefore useless for the reader.} \textcolor{blue}{Corey - Agreed. I'll replace with a table, which should be clearer.}}
%      \label{fig:architectures}
%      \vspace{-0.3cm}
%  \end{figure}

\begin{table}[!t]
\centering
\caption{Final Architectures for CIFAR-10, VWW, and KWS. Other networks for \ac{VWW} and \ac{KWS} are not listed, as they cannot easily be represented using our macro-architecture.%\textcolor{blue}{It is unclear to me why no baseline networks are listed for VWW and KWS?}
}\label{table:final_networks}
\begin{threeparttable}
\resizebox{0.5\textwidth}{!}{\begin{tabular}{lccccccc} 
\toprule \toprule
\multirow{2}{*}{\textbf{Network}} & \multicolumn{7}{c}{\textbf{Macro-Architecture Parameter}}                                       \\
                                  & \textbf{OC$_0$} & \textbf{KS$_0$} & \textbf{M} & \textbf{R*} & \textbf{B*} & \textbf{CT*} & \textbf{WF*}  \\ \midrule
\multicolumn{8}{c}{\textbf{CIFAR-10}}                                                                                               \\ \midrule
Resnet32                          & 64 & 7 & 3 & (5, 5, 5) & (1, 1, 1) & (B, B, B) & (1, 1, 1)\\
AnalogNAS\_T100                   & 32 & 3 & 1 & (2, ) & (1,) & (C, ) & (2, )\\
AnalogNAS\_T300                   & 32 & 3 & 1 & (3, 3) & (1, 1) & (A, B) & (2, 1)\\
AnalogNAS\_T500                   & 64 & 5 & 1 & (3, ) & (3, ) & (A, ) & (2, )\\
AnalogNAS\_T1M                    & 32 & 5 & 2 & (3, 3) & (2, 2) & (A, A) & (3, 3)\\ \midrule
\multicolumn{8}{c}{\textbf{VWW}}                                                                                                    \\ \midrule
AnalogNAS\_T200                   & 24 & 3 & 3 & (2, 2, 2) & (1, 2, 1) & (B, A, A) & (2, 2, 2)\\
AnalogNAS\_T400                   & 68 & 3 & 2 & (3, 5)  & (2, 1) & (C, C) & (3, 2)\\ \midrule
\multicolumn{8}{c}{\textbf{KWS }}                                                                                                   \\ \midrule
AnalogNAS\_T200                   & 80 & 1 & 1 & (1, ) & (2, ) & (C, ) & (4, )\\
AnalogNAS\_T400                   & 68 & 1 & 2 & (2, 1) & (1, 2) & (B, B) & (3, 3)\\
\bottomrule \bottomrule
\end{tabular}}
\begin{tablenotes}
\item *As depicted in Fig.~\ref{fig:macro_architecture}, thse macro-architecture parameters comprise\\multiple instances.
\end{tablenotes}
\end{threeparttable}
\vspace{-0.5cm}
\end{table}

In Fig.~\ref{fig:results}, the simulated hardware comparison of the three tasks is depicted. Our models outperform \ac{SOTA} architectures with respect to both accuracy and resilience to drift. On CIFAR-10, after training the surrogate model, the search took 17 minutes to run. We categorize the results based on the number of parameters threshold into two distinct groups. We consider edge models with a number of parameters below 1M and above 400k. Below 400K, architectures are suitable for TinyML deployment. The final architecture, T500, is smaller than Resnet32, and achieved +1.86\% better accuracy and a drop of 1.8\% after a month of inference, compared to 5.04\%. %\textcolor{red}{I think it would be useful to motivate why we are looking at this timescale. What are the tmescales of interest overall for edgeAI.} Motivation in section IV.B but need to add for edge ?. 
This model is $\sim86\times$ smaller than Wide Resnet~\cite{wideresnet}, which has 36.5M parameters. Our smallest model, T100, was $1.23\times$ bigger than Resnet-V1, the \ac{SOTA} model bench-marked by MLPerf~\cite{mlperf}. Despite not containing any depth-wise convolutions, Resnet\_V1 is extremely small, with only ~70k parameters. Our model offers a +7.98\% accuracy increase with a 5.14\% drop after a month of drift compared to 10.1\% drop for Resnet\_V1. Besides, our largest model, \textit{AnalogNAS\_1M}, outperforms Wide Resnet with +0.86\% in the 1-day accuracy with a drop of only 1.16\% compared to 6.33\%. In addition, the found models exhibit greater consistency across experiment trials, with an average standard deviation of 0.43 over multiple drift times as opposed to 0.97 for \ac{SOTA} models.

Similar conclusions can be made about \ac{VWW} and \ac{KWS}. In \ac{VWW}, current baselines use a depth-wise separable convolution that incurs a high accuracy drop on analog devices. Compared to AnalogNet-VWW and Micronets-VWW, the current \ac{SOTA} networks for \ac{VWW} in analog and edge devices, our T200 model has similar number of parameters (x1.23 smaller) with a +2.44\% and +5.1\% 1-day accuracy increase respectively. AnalogNAS was able to find more robust and consistent networks with an average \ac{AVM} of 2.63\% and a standard deviation of 0.24. MCUNet~\cite{mcunet} and MobileNet-V1 present the highest \ac{AVM}. This is due to the sole use of depth-wise separable convolutions.%textcolor{red}{do we map the DWconv to analog for these networks? If yes, I am very surprised that the accuracy can hold this well...}.

On \ac{KWS}, the baseline architectures, including DSCNN~\cite{dscnn}, use hybrid networks containing recurrent cells and convolutions. The recurrent part of the model ensures high robustness to noise. While current models are already robust with an average accuracy drop of 4.72\%, our model outperforms tiny \ac{SOTA} models with 96.8\% and an accuracy drop of 2.3\% after a month of drift. Critically, our AnalogNAS models exhibit greater consistency across experiment trials, with an average standard deviation of 0.17 over multiple drift times as opposed to 0.36 for \ac{SOTA} models.

\begin{figure}[!b]
    \centering
    \includegraphics[width=0.4\textwidth]{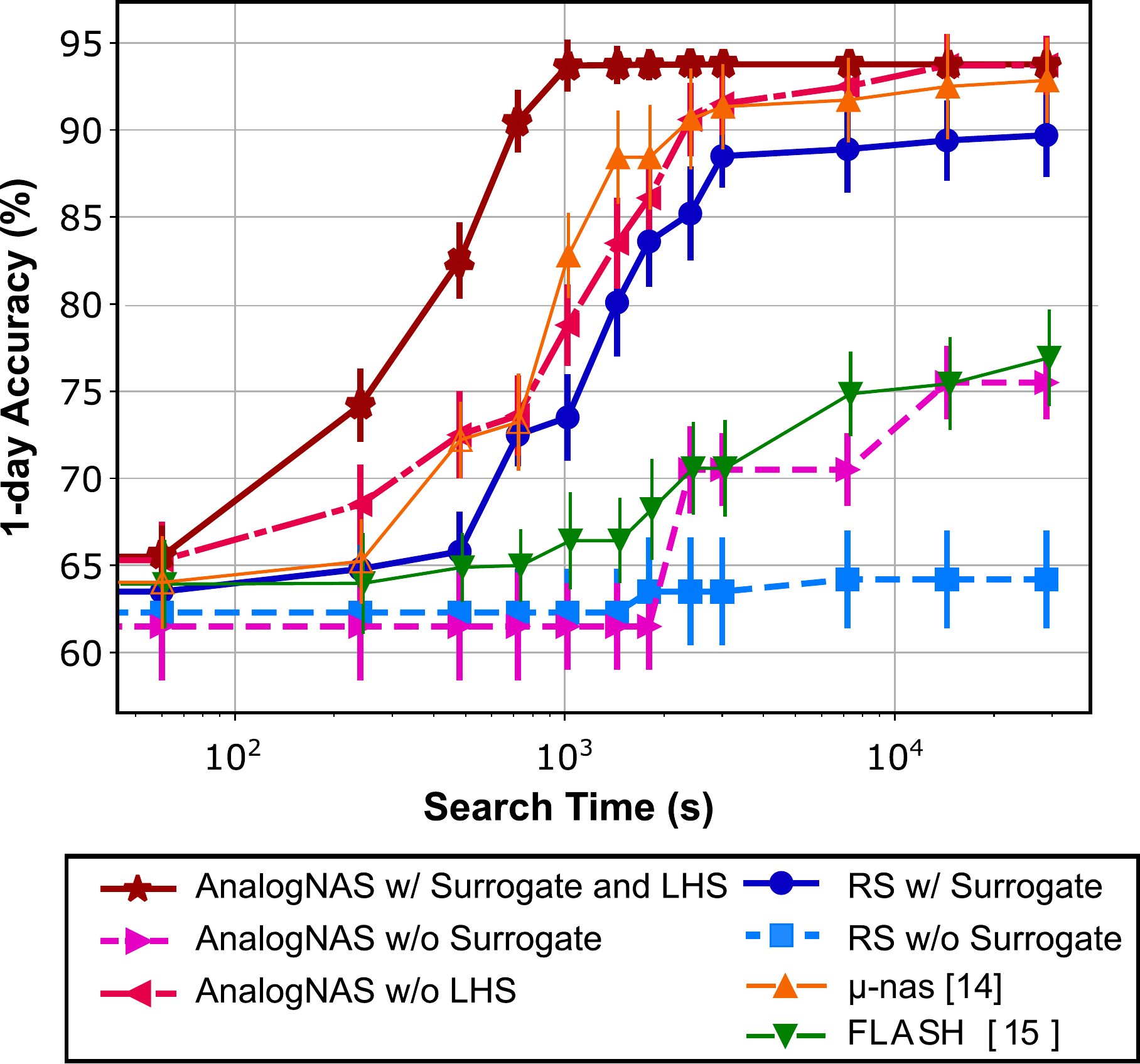}
    \caption{Ablation study comparison against HW-NAS. Mean and standard deviation values are reported across five experiment instances (trials).}
    \label{fig:comp}
\end{figure}

\subsection{Comparison with HW-NAS}
In accordance with commonly accepted \ac{NAS} methodologies, we conducted a comparative analysis of our search approach with Random Search. Results, presented in Fig.~\ref{fig:comp}, were obtained across five experiment instances. Our findings indicate that Random Search was unable to match the 1-day accuracy levels of our final models, even after conducting experiments for a duration of four hours and using the same surrogate model. We further conducted an ablation study to evaluate the effectiveness of our approach by analyzing the impact of the LHS algorithm and surrogate model. The use of a random sampling strategy and exclusion of the surrogate model resulted in a significant increase in search time. The LHS algorithm helped in starting from a diverse initial population and improving exploration efficiency, while the surrogate model played a crucial role in ensuring practical search times.

Besides, AnalogNAS surpasses both FLASH~\cite{flash} and $\mu$ -nas~\cite{micronas} in performance and search time. FLASH search strategy is not adequate for large search spaces such as ours. As for $\mu$-nas, although it manages to achieve acceptable results, its complex optimization algorithm hinders the search process, resulting in decreased efficiency. 

\subsection{Search Time and \acf{AVM} Threshold Trade-Off} 
During the search, we culled architectures using their predicted \ac{AVM}, i.e., any architecture with a higher \ac{AVM} than the \ac{AVM} threshold was disregarded. As listed in Table~\ref{tab:slope}, we varied this threshold to investigate the trade-off between T$_{\text{AVM}}$ and the search time. As can be seen, as T$_{\text{AVM}}$ is decreased, the delta between \ac{AVM} and T$_{\text{AVM}}$ significantly decreases. The correlation between the search time and T$_{\text{AVM}}$ is observed to be non-linear.

% Corey - I'm not quite sure what we are trying to convey with this. I have tentatively removed the following paragraph and re-framed the section as a simple investigation of the trade-off between the search time and the threshold.}

% The architecture obtained with a 1\% threshold is a single convolution block that utilizes the tile at 10\%. The convolution is mapped to a tile in an im2col~\cite{im2col} format, which makes the logical size of the weights to $k_s^2 c_\text{in} \times c_\text{out}$, $k_s$ refers to the kernel size, and $c_\text{in}$ and $c_\text{out}$ are the input and output channel size, respectively. The first convolution has a dimension of $27$, i.e., the RGB depth of three multiplied by the kernel size. Achieving a 10\% utilization in the first convolution requires 255 as the output channel size, which significantly drops the accuracy. Besides, the search time augments whenever we tighten the constraints \textcolor{red}{I find the paragraph very confusing, what is the message? Also, the T500 has 5 percent average tile utilization, so that is also not great...}. 

% \textbf{Insight:} The higher the utilization of the tile is, the more resilient the architecture will be. However, maximizing the tile utilization of all layers significantly impact the accuracy. 

%  \setlength{\arrayrulewidth}{0.5mm}
 \begin{table}[!t]
 \centering
 \caption{\ac{AVM} threshold variation results on CIFAR-10.}\label{tab:slope}
 \begin{threeparttable}
 \begin{tabular}{lrrr}
 \toprule \toprule
 \textbf{T$_{\text{AVM}}$ (\%)} & \textbf{1.0} & \textbf{3.0}  & \textbf{5.0*} \\ \midrule %\hline 
1-day Accuracy             & 88.7\%        &       93.71\%      & 93.71\%                      \\ %\hline
\ac{AVM}       & 0.85\%            & 1.8\%                   & 1.8\%             \\
Search Time (min) & 34.65 & 28.12 & 17.65 \\ 
\bottomrule \bottomrule %\hline
 \end{tabular}
 \begin{tablenotes}
\item *Overall results computed with T$_{\text{AVM}}$ (\%) = 5.0.
\end{tablenotes}
\end{threeparttable}
\vspace{-0.5cm}
 \end{table}

\section{Experimental Hardware Validation and Architecture Performance Simulations}
\label{hardware}

\subsection{Experimental Hardware Validation}
An experimental hardware accuracy validation study was performed using a 64-core \ac{IMC} chip based on \ac{PCM}~\cite{HERMES}. Each core comprises a crossbar array of 256x256 \ac{PCM}-based unit-cells along with a local digital processing unit~\cite{9508706}. This validation study was performed to verify whether the simulated network accuracy values and rankings are representative of those when the networks are deployed on real physical hardware. We deployed two networks for the CIFAR-10 image classification task on hardware: AnalogNAS\_T500 and the baseline ResNet32~\cite{resnet} networks from Fig.~\ref{fig:results}(a).

To implement the aforementioned models on hardware, after \ac{HWA} training was performed, a number of steps were carried out. First, from the \ac{AIHWKIT}, unit weights of linear (dense) and unrolled convolutional layers, were exported to a state dictionary file. This was used to map network parameters to corresponding network layers. Additionally, the computational inference graph of each network was exported. These files were used to generate proprietary data-flows to be executed in-memory. As only hardware accuracy validation was being performed, all other operations aside from \acp{MVM} were performed on a host machine connected to the chip through an \ac{FPGA}.
The measured hardware accuracy was 92.05\% for T500 and 89.87\% for Resnet32, as reported in Table~\ref{table:hardware_efficiency}. Hence, the T500 network performs significantly better than Resnet32 also when implemented on real hardware. This further validates that our proposed AnalogNAS approach is able to find networks with similar number of parameters that are more accurate and robust on analog \ac{IMC} hardware. 

\subsection{Simulated Hardware Energy and Latency}

%We conducted experiments to compare the performance of our final model, AnalogNAS\_T500, with that of ResNet32 model, using architecture performance simulations based off the work presented in Ref.~\cite{ADS}. The platform we simulated was a single analog fabric of 100mm$^2$. Our results, provided in Table~\ref{table:hardware_efficiency}, show that AnalogNAS\_T500 outperformed ResNet32 in terms of both speed and energy efficiency. 
We conducted power performance simulations for AnalogNAS\_T500 and ResNet32 models using a 2D-mesh based heterogeneous analog IMC system with the simulation tool presented in~\cite{ADS}. The simulated IMC system consists of one analog fabric with 48 analog tiles of 512x512 size, on-chip digital processing units, and digital memory for activation orchestration between CNN layers. Unlike the accuracy validation experiments on the 64-core IMC chip, the simulated power performance assumes all intermediate operations to be mapped and executed on-chip. Our results, provided in Table~\ref{table:hardware_efficiency}, show that AnalogNAS\_T500 outperformed ResNet32 in terms of both execution time and energy efficiency.

We believe that this power performance benefit is realized because, in analog \ac{IMC} hardware, wider layers can be computed in parallel, leveraging the $O(1)$ latency from analog tiles, and are therefore preferred over deeper layers. It is noted that both networks exhibit poor tile utilization and that the tile utilization and efficiency of these networks could be further improved by incorporating these metrics as explicit constraints. This is left to future work and is beyond the scope of AnalogNAS.
% TODO:  We did not optimize for tile utilization and this will be the scope of future work.
% Interestingly, we found that the most robust architectures, which were discovered through our search space, were also the most energy efficient. Specifically, we found that wide networks, with larger numbers of output channels and convolutions, were able to better utilize the resources of the analog fabric and achieve better energy efficiency. This observation is supported by the fact that wider networks have more opportunities for parallelism, and can therefore distribute the workload more efficiently across the available analog tiles. 
% \textcolor{red}{I am not sure we can conclude the points that were mentioned here. We only did energy efficiency simulations for one model, we do not have any other data points. Also, the array utilization is very poor in T500. What we can perhaps mention that we get this performance benefit because wider layers are preferred as opposed to deeper. And although we do not have any HW constraints in the search, it still settles to this. We can improve on the HW utilization and perhaps performance/energy benefits if in the future we incorporate these as explicit constraints. }
% our experiments demonstrate the effectiveness of AnalogNAS in discovering efficient and robust neural network architectures for analog hardware. The results suggest that wide networks are a promising avenue for optimizing energy efficiency in analog hardware. \textcolor{red}{b1}

\begin{table}[!t]
\centering
\caption{Experimental hardware accuracy validation and simulated power performance on the IMC system in~\cite{ADS}.}\label{table:hardware_efficiency}
\begin{tabularx}{\columnwidth}{Xrr}
% \toprule \toprule
% \textbf{}  & \textbf{ResNet32}\textcolor{red}{ref}      & \textbf{AnalogNAS\_T500}         \\ \midrule

% \textbf{Hardware accuracy} & 90.19\% & 92.25\% \\\midrule \midrule

% \textbf{Total weights}         & 464,432                & 416,960               \\ 
% \textbf{Total tiles}            & 43                     & 27                     \\ 
% \textbf{Network depth}            & 32                     & 11                     \\ 
% \textbf{Execution time (msec)}           & \textbf{0.433447} & \textbf{0.10816}  \\ 
% %\textbf{Inferences/s}           & 2307.09                & 9245.56               \\ 
% \textbf{Inferences/s/W}          & \textbf{43923.9}       & \textbf{54406.9}      \\
% %\textbf{Average tile utilization} & \textbf{3.35\%}        & \textbf{5.57\%}        \\ 
% \bottomrule \bottomrule
\toprule \toprule
\textbf{Architecture}                         & \textbf{ResNet32} & \textbf{AnalogNAS\_T500}  \\ 
\midrule
\multicolumn{3}{c}{\textbf{Hardware Experiments}}                                                        \\ \midrule
FP Accuracy        & 94.34\%           & 94.54\%       \\
Hardware accuracy*        & 89.55\%           & 92.07\%                   \\ 
\midrule \midrule
\multicolumn{3}{c}{\textbf{Simulated Hardware Power Performance}}                                                           \\ \midrule
Total weights            & 464,432           & 416,960                   \\
Total tiles              & 43                & 27                        \\
Network Depth            & 32                & 17                        \\
Execution Time (msec)    & {0.434} & {0.108}          \\
%\textbf{Inferences/s}             & 2307.09           & 9245.56                   \\
Inferences/s/W           & {43,956.7}  & {54,502}          \\
%\textbf{Average tile utilization} & \textbf{3.35\%}   & \textbf{5.57\%}           \\
\bottomrule \bottomrule
\end{tabularx}
\begin{tablenotes}
\item *The mean accuracy is reported across five experiment repetitions.
\end{tablenotes}
\vspace{-0.5cm}
\end{table}

\section{Discussion}
\label{discussion}
\begin{figure}[!t]
    \centering
    \includegraphics[width=0.45\textwidth]{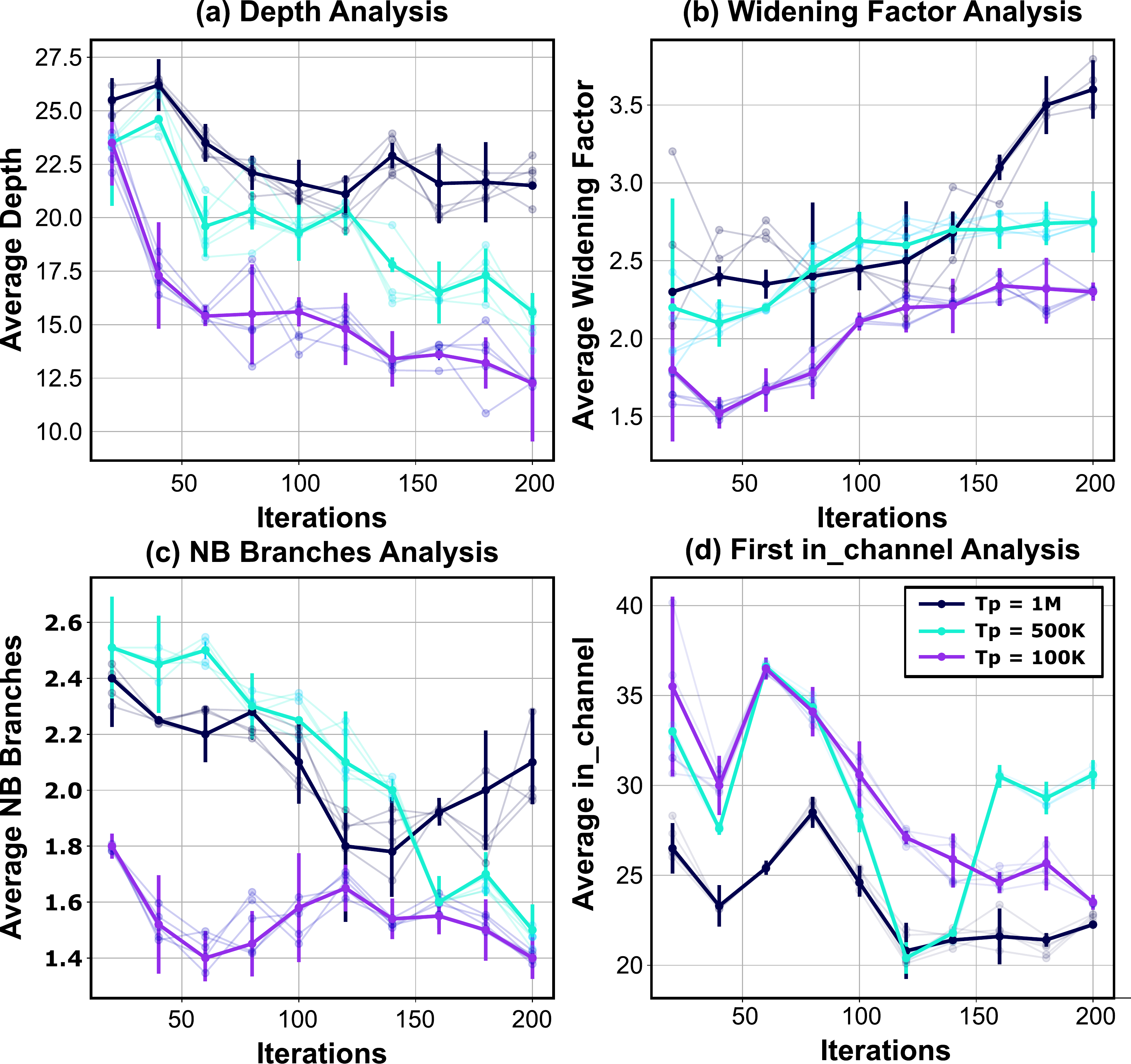}
     \caption{Evolution of architecture characteristics in the population during the search for CIFAR-10. Random individual networks are shown.
     }\label{fig:insights}
\end{figure}

During the search, we analyzed the architecture characteristics and studied which types of architectures perform the best on \ac{IMC} inference processors. The favoured architectures combine robustness to noise and accuracy performance. Fig.~\ref{fig:insights} shows the evolution of the average depth, the average widening factor, the average number of branches, and the average first convolution's output channel size of the search population for every 20 iterations. The depth represents the number of convolutions. % \textcolor{red}{In this paragraph, could we also in parenthesis tie the variables to Fig3 notations? Also, for table IV, I compiled depth as stages, not total num convs. Could you correct this in line with what is presented in Fig9?}. 
A sampled architecture has a widening factor per block. To compute the average widening factor, we first computed the average widening factor per architecture by dividing the sum of the widening factors by the number of blocks contained in the architecture. Then, we calculated the average widening factor across all architectures. Similar computations were performed for the average number of branches. 

For each plot, the search was run 5 times and the mean is represented in each point. The plotted error corresponds to one standard deviation from that mean. 
Starting from a random population obtained using \ac{LHS}, the population evolves through different width and depth-related mutations. During this analysis, we want to answer the following questions: (i) does the search favor wide or deep networks? And subsequently, are wider architectures more noise resilient? (ii) what architectures are exploited by the search for different tasks when constraining the number of parameters? 

\subsection{Are Wider or Deeper Networks More Robust to PCM Device Drift?}
From Fig.~\ref{fig:insights}, it can be observed that the depth of all networks decreases during the search. This trend is especially seen when we constrain the model's size to 100K and 500K parameters. During the search, the widening factor also increases, allowing the blocks to have wider convolutions. The number of branches is highly dependent on $T_p$. This number is, on average, between 1 and 2. The branches are the number of parallel convolutions in a block disregarding the skip connection. In the literature, architectures such as ResNext, that support a higher number of branches, have a number of parameters around 30M. It is still interesting to get blocks with two branches, which also reflects an increase in the width of the network by increasing the number of features extracted within the same block.%\textcolor{red}{How does branch and width correlate?}. 
The average output channel size of the first convolution decreases during the search. Its final value is around the same number of output channels as standard architectures, i.e., between 16 and 32. This follows the general trend of having wider convolutions in deeper network positions. 

\subsection{Types Of Architectures}
The architectures and parameter constraints differ for each task, but they all exhibit an increasing expansion ratio in the convolution block. This allows the convolutions to effectively utilize the tile and mitigate noise from unused cells in the crossbar.%\textcolor{red}{Again, T500 only has 5percent tile utilization...} 
For CIFAR-10, architectures behave like Wide Resnet~\cite{wideresnet} while respecting the number of parameters constraint. For the \ac{VWW} task, the architectures are deeper. The input resolution is $224\times224$, which requires more feature extraction blocks. However, they are still smaller than \ac{SOTA} architectures, with a maximum depth of 22. As depth is essential to obtain high accuracy for the \ac{VWW} task, no additional branches are added. For the \ac{KWS} task, the architectures are the widest possible, maximizing the tile utilization for each convolutional layer.

% \begin{figure*}[!t]
%     \centering
%     \includegraphics[width=0.9\textwidth]{figures/hardware.pdf}
%      \caption{Hardware results.}\label{fig:insights}
% \end{figure*}

\section{Conclusion}\label{conclusion}
In this paper, we propose an efficient \ac{NAS} methodology dedicated to analog in-memory computing for TinyML tasks entitled AnalogNAS. The obtained models are accurate, noise and drift-resilient, and small enough to run on resource-constrained devices. Experimental results demonstrate that our method outperforms \ac{SOTA} models on analog hardware for three tasks of the MLPerf benchmark: image classification on CIFAR-10, \ac{VWW}, and \ac{KWS}. Our AnalogNAS\_T500 model implemented on physical hardware demonstrates $>2\%$ higher accuracy experimentally on the CIFAR-10 benchmark than ResNet32. Calculated speed and energy efficiency estimates reveal a $>4\times$ reduction in execution time, in addition to $>1.2\times$ higher energy efficiency for AnalogNAS\_T500 compared with ResNet32 when evaluated using a system-level simulator.
While our paper has focused on a ResNet-like search space, it is important to note that our search strategy is adaptable and can be extended in future work to explore a broader range of architectures.

\section*{Acknowledgement}
We would like to thank Thanos Vasilopoulos and Julian Büchel  from IBM Research, Zurich lab, for their help with the development of the hardware-software infrastructure. We also thank the IBM Research Analog AI team for their feedback.
We thank the computational support from AiMOS, an AI supercomputer made available by the IBM Research AI Hardware Center and Rensselaer Polytechnic Institute's Center for Computational Innovations (CCI).

\bibliographystyle{IEEEtran}
\bibliography{References}

\begin{thebibliography}{10}
\providecommand{\url}[1]{#1}
\csname url@samestyle\endcsname
\providecommand{\newblock}{\relax}
\providecommand{\bibinfo}[2]{#2}
\providecommand{\BIBentrySTDinterwordspacing}{\spaceskip=0pt\relax}
\providecommand{\BIBentryALTinterwordstretchfactor}{4}
\providecommand{\BIBentryALTinterwordspacing}{\spaceskip=\fontdimen2\font plus
\BIBentryALTinterwordstretchfactor\fontdimen3\font minus
  \fontdimen4\font\relax}
\providecommand{\BIBforeignlanguage}[2]{{%
\expandafter\ifx\csname l@#1\endcsname\relax
\typeout{** WARNING: IEEEtran.bst: No hyphenation pattern has been}%
\typeout{** loaded for the language `#1'. Using the pattern for}%
\typeout{** the default language instead.}%
\else
\language=\csname l@#1\endcsname
\fi
#2}}
\providecommand{\BIBdecl}{\relax}
\BIBdecl

\bibitem{imcgeneral}
A.~Sebastian, M.~Le~Gallo, R.~Khaddam-Aljameh, and E.~Eleftheriou, ``Memory
  devices and applications for in-memory computing,'' \emph{Nature
  Nanotechnology}, vol.~15, no.~7, pp. 529--544, Jul 2020.

\bibitem{DBLP:journals/corr/abs-1906-03138}
V.~Joshi, M.~L. Gallo, I.~Boybat, S.~Haefeli, C.~Piveteau, M.~Dazzi,
  B.~Rajendran, A.~Sebastian, and E.~Eleftheriou, ``Accurate deep neural
  network inference using computational phase-change memory,'' \emph{CoRR},
  vol. abs/1906.03138, 2019.

\bibitem{Lammie2022}
C.~Lammie, W.~Xiang, B.~Linares-Barranco, and M.~{Rahimi Azghadi}, ``Memtorch:
  An open-source simulation framework for memristive deep learning systems,''
  \emph{Neurocomputing}, vol. 485, pp. 124--133, 2022.

\bibitem{BoybatIEDM}
I.~Boybat, B.~Kersting, S.~G. Sarwat, X.~Timoneda, R.~L. Bruce, M.~BrightSky,
  M.~L. Gallo, and A.~Sebastian, ``Temperature sensitivity of analog in-memory
  computing using phase-change memory,'' in \emph{2021 IEEE International
  Electron Devices Meeting (IEDM)}, 2021, pp. 28.3.1--28.3.4.

\bibitem{Howard2017}
A.~G. Howard, M.~Zhu, B.~Chen, D.~Kalenichenko, W.~Wang, T.~Weyand,
  M.~Andreetto, and H.~Adam, ``Mobilenets: Efficient convolutional neural
  networks for mobile vision applications,'' \emph{CoRR}, vol. abs/1704.04861,
  2017.

\bibitem{Lu2022}
G.~Lu, W.~Zhang, and Z.~Wang, ``Optimizing depthwise separable convolution
  operations on gpus,'' \emph{IEEE Transactions on Parallel and Distributed
  Systems}, vol.~33, no.~1, pp. 70--87, 2022.

\bibitem{Benmeziane2021}
H.~Benmeziane, K.~E. Maghraoui, H.~Ouarnoughi, S.~Niar, M.~Wistuba, and
  N.~Wang, ``A comprehensive survey on hardware-aware neural architecture
  search,'' \emph{CoRR}, vol. abs/2101.09336, 2021.

\bibitem{resnext}
S.~Xie, R.~B. Girshick, P.~Doll{\'{a}}r, Z.~Tu, and K.~He, ``Aggregated
  residual transformations for deep neural networks,'' in \emph{{IEEE}
  Conference on Computer Vision and Pattern Recognition, {CVPR}}.\hskip 1em
  plus 0.5em minus 0.4em\relax {IEEE} Computer Society, 2017, pp. 5987--5995.

\bibitem{wideresnet}
S.~Zagoruyko and N.~Komodakis, ``Wide residual networks,'' in \emph{Proceedings
  of the British Machine Vision Conference, {BMVC}}, R.~C. Wilson, E.~R.
  Hancock, and W.~A.~P. Smith, Eds.\hskip 1em plus 0.5em minus 0.4em\relax
  {BMVA} Press, 2016.

\bibitem{nassurvey}
L.~Sekanina, ``Neural architecture search and hardware accelerator co-search:
  {A} survey,'' \emph{{IEEE} Access}, vol.~9, pp. 151\,337--151\,362, 2021.

\bibitem{mlperf}
C.~R. Banbury, V.~J. Reddi, P.~Torelli, N.~Jeffries, C.~Kir{\'{a}}ly,
  J.~Holleman, P.~Montino, D.~Kanter, P.~Warden, D.~Pau, U.~Thakker,
  A.~Torrini, J.~Cordaro, G.~D. Guglielmo, J.~M. Duarte, H.~Tran, N.~Tran,
  W.~Niu, and X.~Xu, ``Mlperf tiny benchmark,'' in \emph{Proceedings of the
  Neural Information Processing Systems Track on Datasets and Benchmarks 1,
  NeurIPS Datasets and Benchmarks 2021, December 2021, virtual}, J.~Vanschoren
  and S.~Yeung, Eds., 2021.

\bibitem{micronet}
C.~R. Banbury, C.~Zhou, I.~Fedorov, R.~M. Navarro, U.~Thakker, D.~Gope, V.~J.
  Reddi, M.~Mattina, and P.~N. Whatmough, ``Micronets: Neural network
  architectures for deploying tinyml applications on commodity
  microcontrollers,'' in \emph{Proceedings of Machine Learning and Systems},
  A.~Smola, A.~Dimakis, and I.~Stoica, Eds.\hskip 1em plus 0.5em minus
  0.4em\relax mlsys.org, 2021.

\bibitem{darts}
\BIBentryALTinterwordspacing
H.~Liu, K.~Simonyan, and Y.~Yang, ``{DARTS:} differentiable architecture
  search,'' \emph{CoRR}, vol. abs/1806.09055, 2018. [Online]. Available:
  \url{http://arxiv.org/abs/1806.09055}
\BIBentrySTDinterwordspacing

\bibitem{micronas}
E.~Liberis, L.~Dudziak, and N.~D. Lane, ``{\(\mu\)}nas: Constrained neural
  architecture search for microcontrollers,'' in \emph{EuroMLSys@EuroSys 2021,
  Proceedings of the 1st Workshop on Machine Learning and Systemsg Virtual
  Event, Edinburgh, Scotland, UK, 26 April, 2021}, E.~Yoneki and P.~Patras,
  Eds.\hskip 1em plus 0.5em minus 0.4em\relax {ACM}, 2021, pp. 70--79.

\bibitem{flash}
G.~Li, S.~K. Mandal, {\"{U}}.~Y. Ogras, and R.~Marculescu, ``{FLASH:} fast
  neural architecture search with hardware optimization,'' \emph{{ACM} Trans.
  Embed. Comput. Syst.}, vol.~20, no.~5s, pp. 63:1--63:26, 2021.

\bibitem{nacim}
W.~Jiang, Q.~Lou, Z.~Yan, L.~Yang, J.~Hu, X.~S. Hu, and Y.~Shi,
  ``Device-circuit-architecture co-exploration for computing-in-memory neural
  accelerators,'' \emph{{IEEE} Trans. Computers}, vol.~70, no.~4, pp. 595--605,
  2021.

\bibitem{nas4rram}
Z.~Yuan, J.~Liu, X.~Li, L.~Yan, H.~Chen, B.~Wu, Y.~Yang, and G.~Sun,
  ``{NAS4RRAM:} neural network architecture search for inference on rram-based
  accelerators,'' \emph{Sci. China Inf. Sci.}, vol.~64, no.~6, 2021.

\bibitem{uncertaintyaware}
Z.~Yan, D.-C. Juan, X.~S. Hu, and Y.~Shi, ``Uncertainty modeling of emerging
  device based computing-in-memory neural accelerators with application to
  neural architecture search,'' in \emph{2021 26th Asia and South Pacific
  Design Automation Conference (ASP-DAC)}, 2021, pp. 859--864.

\bibitem{densenet}
G.~Li, M.~Zhang, J.~Li, F.~Lv, and G.~Tong, ``Efficient densely connected
  convolutional neural networks,'' \emph{Pattern Recognit.}, vol. 109, p.
  107610, 2021.

\bibitem{analognet}
C.~Zhou, F.~Redondo, J.~Buchel, I.~Boybat, X.~Comas, S.~R. Nandakumar, S.~Das,
  A.~Sebastian, M.~L. Gallo, and P.~N. Whatmough, ``Ml-hw co-design of
  noise-robust tinyml models and always-on analog compute-in-memory edge
  accelerator,'' \emph{IEEE Micro}, vol.~42, no.~06, pp. 76--87, 2022.

\bibitem{aihwkit}
M.~J. Rasch, D.~Moreda, T.~Gokmen, M.~L. Gallo, F.~Carta, C.~Goldberg, K.~E.
  Maghraoui, A.~Sebastian, and V.~Narayanan, ``A flexible and fast pytorch
  toolkit for simulating training and inference on analog crossbar arrays,'' in
  \emph{3rd {IEEE} International Conference on Artificial Intelligence Circuits
  and Systems}.\hskip 1em plus 0.5em minus 0.4em\relax {IEEE}, 2021, pp. 1--4.

\bibitem{aihwkit_arxiv}
\BIBentryALTinterwordspacing
M.~J. Rasch, C.~Mackin, M.~L. Gallo, A.~Chen, A.~Fasoli, F.~Odermatt, N.~Li,
  S.~R. Nandakumar, P.~Narayanan, H.~Tsai, G.~W. Burr, A.~Sebastian, and
  V.~Narayanan, ``Hardware-aware training for large-scale and diverse deep
  learning inference workloads using in-memory computing-based accelerators,''
  2023. [Online]. Available: \url{https://arxiv.org/abs/2302.08469}
\BIBentrySTDinterwordspacing

\bibitem{eeea}
C.~Termritthikun, Y.~Jamtsho, J.~Ieamsaard, P.~Muneesawang, and I.~Lee,
  ``Eeea-net: An early exit evolutionary neural architecture search,''
  \emph{Eng. Appl. Artif. Intell.}, vol. 104, p. 104397, 2021.

\bibitem{fbnet}
X.~Dai, A.~Wan, P.~Zhang, B.~Wu, Z.~He, Z.~Wei, K.~Chen, Y.~Tian, M.~Yu,
  P.~Vajda, and J.~E. Gonzalez, ``Fbnetv3: Joint architecture-recipe search
  using predictor pretraining,'' in \emph{{IEEE} Conference on Computer Vision
  and Pattern Recognition, {CVPR}}.\hskip 1em plus 0.5em minus 0.4em\relax
  Computer Vision Foundation / {IEEE}, 2021, pp. 16\,276--16\,285.

\bibitem{DBLP:conf/dac/JiangZSYZSH19}
W.~Jiang, X.~Zhang, E.~H. Sha, L.~Yang, Q.~Zhuge, Y.~Shi, and J.~Hu, ``Accuracy
  vs. efficiency: Achieving both through fpga-implementation aware neural
  architecture search,'' in \emph{Proceedings of the 56th Annual Design
  Automation Conference 2019, {DAC} 2019, Las Vegas, NV, USA, June 02-06,
  2019}.\hskip 1em plus 0.5em minus 0.4em\relax {ACM}, 2019, p.~5.

\bibitem{bayesianhwnas}
A.~Sarah, D.~Cummings, S.~N. Sridhar, S.~Sundaresan, M.~Szankin, T.~Webb, and
  J.~P. Munoz, ``A hardware-aware system for accelerating deep neural network
  optimization,'' \emph{CoRR}, vol. abs/2202.12954, 2022.

\bibitem{hao}
Z.~Dong, Y.~Gao, Q.~Huang, J.~Wawrzynek, H.~K.~H. So, and K.~Keutzer, ``{HAO:}
  hardware-aware neural architecture optimization for efficient inference,'' in
  \emph{29th {IEEE} Annual International Symposium on Field-Programmable Custom
  Computing Machines, {FCCM} 2021, Orlando, FL, USA, May 9-12, 2021}.\hskip 1em
  plus 0.5em minus 0.4em\relax {IEEE}, 2021, pp. 50--59.

\bibitem{efficientnetv2}
M.~Tan and Q.~V. Le, ``Efficientnetv2: Smaller models and faster training,'' in
  \emph{Proceedings of the 38th International Conference on Machine Learning,
  {ICML}}, ser. Proceedings of Machine Learning Research, M.~Meila and
  T.~Zhang, Eds., vol. 139.\hskip 1em plus 0.5em minus 0.4em\relax {PMLR},
  2021, pp. 10\,096--10\,106.

\bibitem{nasnet}
B.~Zoph, V.~Vasudevan, J.~Shlens, and Q.~V. Le, ``Learning transferable
  architectures for scalable image recognition,'' in \emph{2018 {IEEE}
  Conference on Computer Vision and Pattern Recognition, {CVPR} 2018, Salt Lake
  City, UT, USA, June 18-22, 2018}.\hskip 1em plus 0.5em minus 0.4em\relax
  Computer Vision Foundation / {IEEE} Computer Society, 2018, pp. 8697--8710.

\bibitem{resnet}
K.~He, X.~Zhang, S.~Ren, and J.~Sun, ``Deep residual learning for image
  recognition,'' in \emph{{IEEE} Conference on Computer Vision and Pattern
  Recognition, {CVPR}}.\hskip 1em plus 0.5em minus 0.4em\relax {IEEE} Computer
  Society, 2016, pp. 770--778.

\bibitem{lhs}
M.~D. McKay, ``Latin hypercube sampling as a tool in uncertainty analysis of
  computer models,'' in \emph{Proceedings of the 24th Winter Simulation
  Conference, Arlington, VA, USA, December 13-16, 1992}, R.~C. Crain, Ed.\hskip
  1em plus 0.5em minus 0.4em\relax {ACM} Press, 1992, pp. 557--564.

\bibitem{gates}
X.~Ning, Y.~Zheng, T.~Zhao, Y.~Wang, and H.~Yang, ``A generic graph-based
  neural architecture encoding scheme for predictor-based {NAS},'' in
  \emph{Computer Vision - {ECCV} 2020 - 16th European Conference, Glasgow, UK,
  August 23-28, 2020, Proceedings, Part {XIII}}, ser. Lecture Notes in Computer
  Science, A.~Vedaldi, H.~Bischof, T.~Brox, and J.~Frahm, Eds., vol.
  12358.\hskip 1em plus 0.5em minus 0.4em\relax Springer, 2020, pp. 189--204.

\bibitem{hwprnas}
H.~Benmeziane, S.~Niar, H.~Ouarnoughi, and K.~E. Maghraoui, ``Pareto rank
  surrogate model for hardware-aware neural architecture search,'' in
  \emph{International {IEEE} Symposium on Performance Analysis of Systems and
  Software, {ISPASS}}.\hskip 1em plus 0.5em minus 0.4em\relax {IEEE}, 2022, pp.
  267--276.

\bibitem{kendall}
H.~Abdi, ``The kendall rank correlation coefficient,'' \emph{Encyclopedia of
  Measurement and Statistics. Sage, Thousand Oaks, CA}, pp. 508--510, 2007.

\bibitem{cifar10}
A.~Krizhevsky, G.~Hinton \emph{et~al.}, ``Learning multiple layers of features
  from tiny images,'' 2009.

\bibitem{vww}
A.~Chowdhery, P.~Warden, J.~Shlens, A.~Howard, and R.~Rhodes, ``Visual wake
  words dataset,'' \emph{CoRR}, vol. abs/1906.05721, 2019.

\bibitem{resnest}
H.~Zhang, C.~Wu, Z.~Zhang, Y.~Zhu, H.~Lin, Z.~Zhang, Y.~Sun, T.~He, J.~Mueller,
  R.~Manmatha, M.~Li, and A.~J. Smola, ``Resnest: Split-attention networks,''
  in \emph{{IEEE/CVF} Conference on Computer Vision and Pattern Recognition
  Workshops, {CVPR}}.\hskip 1em plus 0.5em minus 0.4em\relax {IEEE}, 2022, pp.
  2735--2745.

\bibitem{rmsprop}
M.~C. Mukkamala and M.~Hein, ``Variants of rmsprop and adagrad with logarithmic
  regret bounds,'' in \emph{Proceedings of the 34th International Conference on
  Machine Learning, {ICML}}, ser. Proceedings of Machine Learning Research,
  D.~Precup and Y.~W. Teh, Eds., vol.~70.\hskip 1em plus 0.5em minus
  0.4em\relax {PMLR}, 2017, pp. 2545--2553.

\bibitem{speechcommandsv2}
\BIBentryALTinterwordspacing
P.~{Warden}, ``{Speech Commands: A Dataset for Limited-Vocabulary Speech
  Recognition},'' \emph{ArXiv e-prints}, Apr. 2018. [Online]. Available:
  \url{https://arxiv.org/abs/1804.03209}
\BIBentrySTDinterwordspacing

\bibitem{melfreq}
J.~Mart{\'{\i}}nez, H.~P{\'{e}}rez{-}Meana, E.~E. Hern{\'{a}}ndez, and M.~M.
  Suzuki, ``Speaker recognition using mel frequency cepstral coefficients
  {(MFCC)} and vector quantization {(VQ)} techniques,'' in \emph{22nd
  International Conference on Electrical Communications and Computers,
  {CONIELECOMP}}, P.~B. S{\'{a}}nchez, R.~Rosas{-}Romero, and M.~J.~O. Galindo,
  Eds.\hskip 1em plus 0.5em minus 0.4em\relax {IEEE}, 2012, pp. 248--251.

\bibitem{adam}
D.~P. Kingma and J.~Ba, ``Adam: {A} method for stochastic optimization,'' in
  \emph{3rd International Conference on Learning Representations}, Y.~Bengio
  and Y.~LeCun, Eds., 2015.

\bibitem{mcunet}
J.~Lin, W.~Chen, Y.~Lin, J.~Cohn, C.~Gan, and S.~Han, ``Mcunet: Tiny deep
  learning on iot devices,'' in \emph{Advances in Neural Information Processing
  Systems 33: Annual Conference on Neural Information Processing Systems},
  H.~Larochelle, M.~Ranzato, R.~Hadsell, M.~Balcan, and H.~Lin, Eds., 2020.

\bibitem{dscnn}
P.~M. S{\o}rensen, B.~Epp, and T.~May, ``A depthwise separable convolutional
  neural network for keyword spotting on an embedded system,'' \emph{{EURASIP}
  J. Audio Speech Music. Process.}, vol. 2020, no.~1, p.~10, 2020.

\bibitem{HERMES}
\BIBentryALTinterwordspacing
M.~L. Gallo, R.~Khaddam-Aljameh, M.~Stanisavljevic, A.~Vasilopoulos,
  B.~Kersting, M.~Dazzi, G.~Karunaratne, M.~Braendli, A.~Singh, S.~M. Mueller,
  J.~Buechel, X.~Timoneda, V.~Joshi, U.~Egger, A.~Garofalo, A.~Petropoulos,
  T.~Antonakopoulos, K.~Brew, S.~Choi, I.~Ok, T.~Philip, V.~Chan, C.~Silvestre,
  I.~Ahsan, N.~Saulnier, V.~Narayanan, P.~A. Francese, E.~Eleftheriou, and
  A.~Sebastian, ``A 64-core mixed-signal in-memory compute chip based on
  phase-change memory for deep neural network inference,'' 2022. [Online].
  Available: \url{https://arxiv.org/abs/2212.02872}
\BIBentrySTDinterwordspacing

\bibitem{9508706}
R.~Khaddam-Aljameh, M.~Stanisavljevic, J.~F. Mas, G.~Karunaratne, M.~Braendli,
  F.~Liu, A.~Singh, S.~M. Müller, U.~Egger, A.~Petropoulos, T.~Antonakopoulos,
  K.~Brew, S.~Choi, I.~Ok, F.~L. Lie, N.~Saulnier, V.~Chan, I.~Ahsan,
  V.~Narayanan, S.~R. Nandakumar, M.~L. Gallo, P.~A. Francese, A.~Sebastian,
  and E.~Eleftheriou, ``Hermes core – a 14nm cmos and pcm-based in-memory
  compute core using an array of 300ps/lsb linearized cco-based adcs and local
  digital processing,'' in \emph{2021 Symposium on VLSI Technology}, 2021, pp.
  1--2.

\bibitem{ADS}
S.~Jain, H.~Tsai, C.-T. Chen, R.~Muralidhar, I.~Boybat, M.~M. Frank,
  S.~Woźniak, M.~Stanisavljevic, P.~Adusumilli, P.~Narayanan, K.~Hosokawa,
  M.~Ishii, A.~Kumar, V.~Narayanan, and G.~W. Burr, ``A heterogeneous and
  programmable compute-in-memory accelerator architecture for analog-ai using
  dense 2-d mesh,'' \emph{IEEE Transactions on Very Large Scale Integration
  (VLSI) Systems}, vol.~31, no.~1, pp. 114--127, 2023.

\end{thebibliography}

\end{document}